\documentclass[aps,12pt,english,tightenlinesletterpaper,superscriptaddress,secnumarabic,nofootinbib]{revtex4}
\DeclareRobustCommand{\baselinestretch{1.0}}

\usepackage[letterpaper]{geometry}
\usepackage{amsmath}
\usepackage{graphicx}
\usepackage{amssymb}
\usepackage{epsfig}
\usepackage{epstopdf}
\usepackage{slashed}

\newcommand{\SU}[1]{\mathrm{SU}\bigl(#1\bigr)}
\newcommand{\SO}[1]{\mathrm{SO}\bigl(#1\bigr)}
\newcommand{\U}[1]{\mathrm{U}\bigl(#1\bigr)}
\newcommand{\snuc}{\tilde{\nu}^c}
\newcommand{\snu}{\tilde{\nu}}
\newcommand{\ch}{\mathrm{ch}}
\newcommand{\eff}{\mathrm{eff}}
\newcommand{\sph}{\mathrm{sph}}
\newcommand{\hc}{\mathrm{h.c.}}
\newcommand{\GeV}{\mathrm{\ GeV}}
\newcommand{\TeV}{\mathrm{\ TeV}}
\newcommand{\ord}[1]{\mathcal{O} \left( #1 \right)}
\newcommand{\eref}[1]{\ensuremath{\mathrm{Eq.}\;(\ref{#1})}}
\newcommand{\y}{\vec{y}}
\newcommand{\x}{\vec{x}}

\usepackage{babel}

\usepackage{hyperref} 
\hypersetup{
    colorlinks=false,       
    linkcolor=red,          
    citecolor=green,        
    filecolor=magenta,      
    urlcolor=cyan           
}

\begin{document}

\title{Electroweak Phase Transition in the $\mu\nu$SSM}

\date{09/22/10}

\author{Daniel J.~H.~Chung}
\email{danielchung@wisc.edu}
\affiliation{
{\small University of Wisconsin-Madison, Department of Physics} \\
{\small 1150 University Avenue, Madison, WI 53706, USA}}

\affiliation{
{\small School of Physics, Korea Institute for Advanced Study, 207-43,}\\
{\small Cheongnyangni2-dong, Dongdaemun-gu, Seoul 130-722, Korea}}

\author{Andrew J. Long}
\email{ajlong@wisc.edu}
\affiliation{
{\small University of Wisconsin-Madison, Department of Physics} \\
{\small 1150 University Avenue, Madison, WI 53706, USA}}

\begin{abstract}
An extension of the MSSM called the $\mu\nu$SSM does not allow a conventional
thermal leptogenesis scenario because of the low scale seesaw that it
utilizes.  Hence, we investigate the possibility of electroweak
baryogenesis.  Specifically, we identify a parameter region for which
the electroweak phase transition is sufficiently strongly first order
to realize electroweak baryogenesis.  In addition to transitions that
are similar to those in the NMSSM, we find a novel class of phase
transitions in which there is a rotation in the singlet vector space.
\end{abstract}
\maketitle

\section{Introduction}

An extension of the MSSM called the $\mu\nu$SSM \cite{LopezFogliani:2005yw} is
a model similar to the NMSSM \cite{Ellis:1988er} (with the usual
$\mathbb{Z}_{3}$ charge assignment) except that the singlet whose
vacuum expectation value (VEV) gives rise to the $\mu$ term also
serves the role of a right-handed neutrino, thereby violating
R-parity. Because the VEV generates the $\mu$-term and the right
handed neutrino mass, the right-handed neutrino masses are of order
TeV, leading to a low scale type I seesaw. Given the absence of a high scale
seesaw, thermal leptogenesis is difficult in the $\mu\nu$SSM.  Hence, it
is interesting to consider whether or not electroweak baryogenesis (EWBG)
\cite{Kuzmin:1985mm} can occur in this class of models.   One of the
most stringent constraints of EWBG on the $\mu\nu$SSM is the requirement
of a sufficiently strongly first order phase transition (SFOPT) such
that the created baryons are not washed out \cite{Bochkarev:1990gb}.

Because the $\mu\nu$SSM contains 3 singlet chiral superfields (right
handed neutrinos), mainly motivated by generality, standard model
generation replication pattern, and phenomenological convenience
\cite{LopezFogliani:2005yw,Escudero:2008jg}, there is a ``larger''
SFOPT parameter space for EWBG when compared to the NMSSM.  More
precisely, there can be SFOPT where the singlet VEVs rotate in the
singlet vector space during the electroweak phase transition.  The
price paid for this is a more complicated global minimum analysis at
both finite and zero temperatures.  The aim of this paper is not to
uncover the most general parameter space consistent with EWBG, but is
to simply give a couple of parametric regions to show the existence of
possibilities.

Depending on the path of the phase transition, the exact $\mu\nu$SSM
parametric dependence of the phase transition strength $v(T_c)/T_c$ is
complicated.  Nonetheless, we find that it is typically true that to
achieve SFOPT, the parameters are close to satisfying the following
condition:
\begin{equation}
\frac{E_{\eff}}{\lambda_{\eff} v(0)} \approx \frac{1}{2}
\end{equation}
where $E_{\eff}$ is the effective cubic coupling,
$\lambda_{\eff}$ is the effective quartic coupling, and $v(0)$
is the magnitude of the scalar field space VEV (including both the
Higgs and singlets) at zero temperature.  Physically, this corresponds
to the parametric region where the critical temperature $T_c$ is small
compared to $v(0)$ during the electroweak phase transition.  In the
examples provided in this paper, whether or not the SFOPT proceeds
from the origin, the leading nonvanishing value of $E_{\eff}$ in the $\mu\nu$SSM
arises from the soft terms
\begin{equation}
\sum_{i}^{3}\left[-a_{\lambda}H_{1}H_{2}\tilde{\nu}_{i}^{c}+\frac{1}{3}a_{\kappa}\left(\tilde{\nu}_{i}^{c}\right)^{3}+\hc \right]
\end{equation} 
where $\tilde{\nu}_{i}^{c}$ are singlet fields.  The dimensionful
coupling $a_\lambda$ is distinguished from $a_\kappa$ in that
$a_\lambda$ also enhances the mixing between the Higgs sector and the
singlet sector.  The leading contribution to $\lambda_{\eff}$ comes from the
superpotential and D-terms.

Beyond these general results, we find a somewhat interesting feature
because we focus on the parametric region analyzed by
\cite{Escudero:2008jg}.  In this parametric region, an approximate
$\mathbb{S}_3$ symmetry (permutation symmetry) arises due to the right
handed neutrino generation independence of the non-Yukawa couplings
and the smallness of the neutrino Yukawa couplings.  Hence, to avoid
any extra complications associated with domain wall formations, one
might naively try to avoid $\mathbb{S}_3$ symmetry breaking phase
transitions by considering parameters which yield zero temperature
vacua preserving $\mathbb{S}_3$.  Hence, this is the boundary
condition that we impose in this paper.  Interestingly, we find that
despite this boundary condition, $\mathbb{S}_3$ is typically
spontaneously broken multiply at finite temperatures in a way that is
sensitive to quantum radiative corrections.  As the temperature is
lowered from high temperatures, this leads to multistep phase
transitions starting from the trivially $\mathbb{S}_3$ symmetric
vacuum in which all VEVs vanish. The electroweak symmetry breaking
phase transition occurs with $\mathbb{S}_3$ symmetry {\em restoration}
to a vacuum in which all sneutrino VEVs are identical and
nonvanishing.  We also find one step SFOPTs in which the scalar fields
(including the singlet fields) make a transition from the origin to
the electroweak symmetry breaking minimum.  The numerical values of
the parametric regions uncovered in this paper is in the paragraph
containing Eq.~(\ref{eq:sigmadef}) and regions IIIa and IIIb depicted
in Fig.~\ref{fig:Scan1}.

There have been many studies of EWBG and the electroweak phase transition 
in models with gauge singlets; some of these are 
Refs.~\cite{McDonald:1993ey, Ham:2004nv, Huber:2006wf, Ham:2007xz, Ham:2007wu, Ham:2007wc, Chiang:2009fs, Kang:2009rd, Kang:2004pp}, and we will discuss others throughout the remainder of this section.  
Since our work is most closely related to previous work on SFOPT in
the NMSSM, we give here a little preview of some of the differences
between our work and select previous works, in addition to the
multidimensional aspect stressed above. 
In Ref.~\cite{Pietroni:1992in}, SFOPT in the context of the NMSSM is
first analyzed and the author points out that the tree level cubic
term coming from the soft SUSY-breaking sector is important. Note that
Ref.~\cite{Pietroni:1992in} uses the definition of critical temperature in
which scalar mass squared matrix develops a vanishing eigenvalue. We
take a more robust definition of $T_{c}$ being the temperature at
which a new coexistence phase occurs even though this definition is
harder to implement in practice.

The authors of Ref.~\cite{Davies:1996qn} also analyze the NMSSM, but
they include a $\mu$-term on the basis that it is more general and its
nonzero value eliminates the $\mathbb{Z}_{3}$ symmetry which can be
cosmologically dangerous with respect to the problem of domain wall
formation \cite{Abel:1995wk}.  The nonzero $\mu$-term leads to false
vacuum not being at the origin.  In this case the critical temperature
criterion used by Ref.~\cite{Pietroni:1992in} is invalid. Therefore,
the authors of Ref.~\cite{Davies:1996qn} take the coexistence phase
definition of critical temperature as we do in this paper.  They also
include a bilinear soft term in the Higgs which breaks the
$\mathbb{Z}_{3}$ symmetry.  Although we do not include such
$\mathbb{Z}_{3}$ breaking terms directly, we will assume that
nonrenormalizable terms can be included to obtain acceptable
phenomenology with respect to any possible domain wall formation.
However, it is to be noted that $\mathbb{Z}_3$ breaking can often
lead to UV instabilities in the singlet tadpoles, making the UV
stability of these theories (including the one considered in this
paper) a model building challenge as noted by \cite{Abel:1995wk}.

The analysis \cite{Huber:2000mg} considers the generalized NMSSM
similar to \cite{Davies:1996qn}. They run 9 parameters with a popular
choice of ``universal'' boundary conditions from the GUT scale down to
the electroweak scale to generate their model. They do not reject
metastable vacua based on the intuition that longevity of the false
vacuum on the horizon scale today is not difficult to attain.  To be
conservative and to avoid potentially complicated discussions of
metastability, we accept only stable vacua in this paper. 

A model related to the NMSSM and the $\mu\nu$SSM is the nMSSM in which
the discrete charge assignment is modified as to eliminate the singlet
cubic term in the superpotential. This model was analyzed by
\cite{Menon:2004wv} for SFOPT. For a significant portion of the
parameter space in which SFOPT occurs, a linear tadpole term in the
superpotential plays a significant role in contrast to our scenario.


The analysis of \cite{Profumo:2007wc} considers the EWPT in an
extension of the SM which adds a real singlet $S$.  These authors find
a large region of the parameter space of their model that is
consistent with SFOPT and LEP Higgs search bounds.  They argue that the
strength of the phase transition can be enhanced by 1) choosing a
large negative value for the $S H^2$ coupling, 2) choosing a negative
value for the $S^2 H^2$ coupling, or 3) allowing the singlet to have a
nonzero VEV before the electroweak symmetry is broken.  In the
language of this paper, the first two points correspond to increasing
$E_{\mathrm{eff}}$ and decreasing $\lambda_{\mathrm{eff}}$,
respectively.  

Before we begin the main body of the work, let us list here all the
caveats to our analysis.  We do not take into account explicitly the
high energy Landau pole constraint (i.e. perturbativity up to the GUT
scale) because we will take the attitude that the $\mu\nu$SSM is well
motivated mainly by its ability to have all fields participate at low
energy and thereby have potential measurability.
Nonetheless, the parametric region that we uncover lies at the border
of perturbativity up to the GUT scale (inferring from the work of
Refs.~\cite{Escudero:2008jg,Miller:2003ay}), which means that the UV
cutoff for our theory can be taken to be far above the TeV scale.  We
do not take into account explicit $\mathbb{Z}_{3}$ breaking effects
because a small amount of breaking can address most cosmological domain
wall problems, as we later demonstrate.  We do not take into account
explicit CP violation effects in the phase transitions as this will
typically lead to less than order 10\% effects since CP violating
phases compatible with phenomenology are typically order $0.1$ or
smaller.  For robustness, we accept in this paper as phenomenological
possibility only absolutely stable global zero temperature vacua
instead of analyzing the phenomenological possibilities of metastable
vacua.  Finally, all of our numerical work is kept in control to only
order 10\% accuracy.

The order of presentation is as follows.  In the next section, we
present the Lagrangian including its discrete symmetry properties and
radiative/thermal corrections.  The section concludes by highlighting
the $\mu\nu$SSM differences from the NMSSM scenario.  In
Sec.~\ref{sec:qualitativedesc}, we describe the parametric region
relevant for SFOPT in terms of one-dimensional field space slice
parametrization.  There we also qualitatively describe how the
multidimensional paths of the phase transition and discrete
symmetries play a role.  Next, in Sec.~\ref{sec:sphaleron}, we
explicitly show that singlets do not play a significant role in terms
of numerical value of the sphaleron action controlling the $B+L$
violating rate in the broken phase.  The main numerical results are
presented in Sec.~\ref{sec:paramscan} where the explicit existence of
the SFOPT parameter region is demonstrated.  Details of the transition
paths organized in terms of discrete symmetries, phenomenological
bounds placed, and explicit mass spectra for a sample parametric point
are given.  In Sec.~\ref{sec:domainwall}, we demonstrate that the 
cosmological domain wall problem is easily evaded with an inclusion of
a weak $\mathbb{Z}_3$ symmetry breaking operator in our scenario.  We
then conclude with a summary of the results.  Several appendices then
follow giving useful technical details.  In Appendix A, we list the
field-dependent mass matrices used for computing the effective
potential.  In the next appendix, we give details regarding the
approximate thermal masses used in the paper.  In Appendix C, we
describe analytically the boundaries of in Fig. \ref{fig:Scan1}
which is one of our main results.  Finally, in Appendix D, we show
that it is generically possible to construct a nonrenormalizable
$\mathbb{Z}_3$ superpotential to obtain a CP conserving global
minimum in the absence of any explicit CP violating parameters.

\section{\label{sec:thermalpotdiff} The Thermal Potential Differences Between the NMSSM and the $\mu\nu$SSM}

The $\mu\nu$SSM that we consider in this paper is specified by the
following superpotential and soft terms
\begin{equation}
W=\sum_{i}^{3} \left\{ Y_{u}^{i} \hat{Q}_{i} \! \cdot \!  \hat{H}_{2} \hat{u}_{i}^{c}+Y_{d}^{i} \hat{H}_{1} \! \cdot \! \hat{Q}_{i} \hat{d}_{i}^c+Y_{e}^{i} \hat{H}_{1} \! \cdot \! \hat{L}_{i}  \hat{e}_{i}^{c} +Y_{\nu}^{i}  \hat{L}_{i} \! \cdot \! \hat{H}_{2} \hat{\nu}_{i}^c-\lambda \hat{H}_{1} \! \cdot \!  \hat{H}_{2} \hat{\nu}_{i}^{c}+\frac{1}{3}\kappa\left( \hat{\nu}_{i}^{c}\right)^{3}\right\} \label{eq:superpotential}\end{equation}
\begin{eqnarray}
-\mathcal{L}_{\mathrm{soft}} & = & \sum_{i}^{3}\left\{
m_{\tilde{Q}}^{2}|\tilde{Q}_{i}|^{2}+m_{\tilde{u}^{c}}^{2}|\tilde{u}_{i}^{c}|^{2}+m_{\tilde{d}^{c}}^{2}|\tilde{d}_{i}^{c}|^{2}+m_{\tilde{L}}^{2}|\tilde{L_{i}}|^{2}+m_{\tilde{e}^{c}}^{2}|\tilde{e}_{i}^{c}|^{2}+m_{\tilde{\nu}^{c}}^{2}|\tilde{\nu}_{i}^{c}|^{2}\right\} \nonumber \\ 
&&+\sum_{i}^{2}m_{H_{i}}^{2}|H_{i}|^{2}-\frac{1}{2}\left(\sum_{i}^{3}M_{i}\tilde{\lambda}_{i}\tilde{\lambda}_{i}+\hc \right)+\sum_{i}^{3}\left[-a_{\lambda}H_{1} \! \cdot \! H_{2}\tilde{\nu}_{i}^{c}+\frac{1}{3}a_{\kappa}\left(\tilde{\nu}_{i}^{c}\right)^{3}+\hc \right] \nonumber \\ 
& & +\sum_{i}^{3}\left\{
a_{u} \tilde{Q}_{i} \! \cdot \! H_{2} \tilde{u}_{i}^{c}+a_{d} H_{1} \! \cdot \! \tilde{Q}_{i} \tilde{d}_{i}^{c}+a_{e} H_{1} \! \cdot \! \tilde{L}_{i} \tilde{e}_{i}^{c}+a_{\nu} \tilde{L}_{i} \! \cdot \! H_{2} \tilde{\nu}_{i}^{c}+\hc \right\}  .
\label{eq:softterms} 
 \end{eqnarray} 
Where indicated by a dot, the $\SU{2}$ indices are contracted with the
antisymmetric tensor and $\epsilon_{12} = 1$.  First, note in addition
to the usual $\mathbb{Z}_3$ symmetry used to forbid an explicit 
$\mu$-term, there is an exact CP symmetry due to the reality of the
coupling constants.  We ignore the CKM phases since these will only
give corrections smaller than the $\mathcal{O}(10\%)$ accuracy that we
are aiming for in this paper.  The CP transformation in the scalar 
effective potential effectively takes each scalar field
to its complex conjugate.  Next, note that the couplings of the 
$\snuc_i$ sector to the SM were taken to be generation independent,
except for the Yukawa couplings, and that the singlets do not couple
with one another directly in the superpotential.  This choice is
motivated by trying partially to match the work of
\cite{Escudero:2008jg}.  Hence, we see there is an enhanced
$\mathbb{S}_3$ symmetry (permutation symmetry) in the
$\tilde{\nu}^c_{i}$ sector if we neglect the Yukawa couplings.  This
approximate symmetry $\mathbb{S}_3$ is nearly exact because of the
smallness of the symmetry breaking Yukawa couplings $Y_{e,\nu}^i$.
As discussed in the Introduction, the exact global $\mathbb{Z}_3$
symmetry itself is plausibly assumed to be broken by nonrenormalizable
operators such that a cosmological domain wall problem does not arise.

At tree level there is an additional global symmetry in the phase in
which the electroweak symmetry is unbroken where $H_i=0$ and all
electromagnetically charged scalars vanish.  Hence, in the high
temperature phase in which the nonsinglet fields are assumed to be
frozen at their classical potential minimum, we have an enhanced
symmetry in the effective potential as a function of the singlets
only.  The enhanced tree level symmetry is $\mathbb{Z}_3 \otimes
\mathbb{Z}_3 \otimes \mathbb{Z}_3$, where each singlet can be phase
rotated independently:
\begin{equation}
\tilde{\nu}_j^c \longrightarrow e^{i n_j 2\pi/3} \tilde{\nu}_j^c.
\label{eq:Z3independent}
\end{equation}
This symmetry appears because we have tuned the superpotential
$\snuc_1 \snuc_2 \snuc_3$ coupling to vanish.  Unlike the approximate
$\mathbb{S}_3$ symmetry, this high-temperature phase classical
symmetry has significant breaking at 1-loop from perturbative
interactions even about the electroweak symmetry preserving minima.
Nonetheless, this $\left( \mathbb{Z}_3 \right)^3$ will be useful in 
understanding the SFOPT in which
there is a rotation in the singlet sector space during the phase
transition.\footnote{One may also wonder whether omitting the
  $\tilde{\nu_1^c} \tilde{\nu_2^c} \tilde{\nu_3^c}$ term is
  radiatively stable.  It turns out that this term is generated at
  2-loop order, which means that as far as one-loop analysis of this
  paper is concerned, this term can be omitted self-consistently.
  However, this must be viewed as fine tuning motivated by staying
  consistent with Ref.~\cite{Escudero:2008jg}.}

In addition to the Yukawa and gauge couplings, there are 19 adjustable
parameters in this model which are
$\{\lambda,\kappa,m_{\tilde{Q},\tilde{u}^{c},\tilde{d}^{c},\tilde{L},\tilde{e}^{c},\tilde{\nu}^{c}}^{2},m_{H_{1,2}}^{2},M_{i},a_{u,d,e,\nu,\lambda,\kappa}\}$.
Because the neutrino Yukawa couplings control the neutrino Dirac mass via the up-type Higgs VEV, these Yukawas are small for reasonable values of $\tan \beta$, 
\begin{equation}
Y_{\nu} \approx 6 \times 10^{-7} \left(
\frac{\sin[\beta]}{\sin[\arctan 2.6]} \right)^{-1},
\end{equation}
and will play a negligible dynamical role. 
The neutral scalar components belonging to the fields
$\{H_{1},H_{2},\tilde{L}_{i},\tilde{\nu}^{c}_i\}$ have finite temperature 
VEVs denoted by 
\begin{equation}
 \left<  H_{i}^{0} \right>_T =v_{i}(T), \,\,\,\,\,\,\,\,\left< \tilde{\nu}_{j}^{(c)} \right>_T =v_{\tilde{\nu}_{j}^{(c)}}(T) \,\,\,\,\,\,\,\,j\in\{1,2,3\}
\end{equation} 
where the braces represent evaluating the field at the global minimum of the 
thermal effective potential \eref{eq:VeffT}.  
To maintain the $\mathbb{S}_3$ symmetry in the electroweak symmetry
breaking vacuum at zero temperature and to avoid unnecessary
complexities in the $SU(2)_L$ charged sector, we choose the
sneutrino VEVs to be independent of generation, such that
\begin{equation}
  \{v_{i}(0)=\mbox{real},v_{\tilde{\nu}_{i}}(0)=v_{\tilde{\nu}}(0)=\mbox{real},v_{\tilde{\nu}_{i}^{c}}(0)=v_{\tilde{\nu}^{c}}(0)=\mbox{real}\}
  .  \label{eq:simplifiedvevatzeroT}
  \end{equation}
To fix the VEVs in this way, we solve the potential minimization condition for the four parameters $\{m_{H_{1}}^{2},m_{H_{2}}^{2},m_{L}^{2},m_{\tilde{\nu}^{c}}^{2}\}$.  
In addition to the
desire to simplify the phase transition history, one of our main
motivations in choosing
$v_{\tilde{\nu}_{i}^{c}}(0)=v_{\tilde{\nu}^{c}}(0)$ is to preserve the
$\mathbb{S}_3$ symmetry manifest in Eqs.~(\ref{eq:superpotential}) and
(\ref{eq:softterms}).  Interestingly enough, as we will see, this
$\mathbb{S}_3$ symmetry spontaneously breaks at finite temperature.  The left-handed sneutrino VEV is small, as we argue below, and the Higgs VEVs satisfy
$v_{1}^{2}(0)+v_{2}^{2}(0) + 3 v_{\snu}^2 (0) \approx v_{1}^{2}(0)+v_{2}^{2}(0)  =v^{2}(0)=(174\mbox{ GeV})^{2}.$  The rest of the parameter specification will be discussed in Sec.~\ref{sec:paramscan}.

To study the electroweak phase transition we need to calculate the
thermal effective potential $V_{\eff}^T$ as a function of temperature
and the field directions which participate in the phase transition.
There are no charged scalars with VEVs at zero temperature and we
assume that there are no charge or color breaking minima to appear at
finite temperature.  In general, the left-handed sneutrinos receive
VEVs $v_{\snu} (0)$ and participate in electroweak symmetry breaking,
but these VEVs must be much less than the electroweak scale to avoid
excessive stellar energy loss by $\tilde{\nu}$
emission \cite{Masiero:1990uj}.  Hence, we can neglect $\ord{v_{\snu}}$ 
contributions and
reduce the relevant degrees of freedom to the five dimensional complex
field space $\left\{ H_1^0, H_2^0, \snuc_{i} \right\}$.
Although a part of the complex phase degrees of freedom in the Higgs
sector is a gauge degree of freedom, for simplicity we will use the
notation $\left\{ H_1^0, H_2^0, \snuc_{i} \right\}$ and keep in mind
that there are nine real degrees of freedom.

We compute the zero temperature effective potential as a loop expansion over the field space $\left\{ H_1^0, H_2^0, \tilde{\nu}_i^c \right\}$.  The leading order term is the tree level potential given by 
\begin{align}\label{eq:V0}
	V_0 
	& = m_{H_1}^2 \left| H_1^0 \right|^2 + m_{H_2}^2 \left| H_2^0 \right|^2 
	+ m_{\tilde{\nu}^c}^2 \sum_i \left| \tilde{\nu}_i^c \right|^2 + \frac{g_1^2 + g_2^2}{8} \left( \left| H_1^0 \right|^2 - \left| H_2^0 \right|^2 \right)^2 \\
	&+ \sum_{i} \left[-a_{\lambda}H_{1}^0 H_{2}^0 \tilde{\nu}_{i}^{c}+\frac{1}{3}a_{\kappa}\left(\tilde{\nu}_{i}^{c}\right)^{3} - \kappa \lambda \left( H_1^0 H_2^0 \right)^{\star}  \left( \snuc_i \right)^2 +\hc \right]  \nonumber \\
	& + 3 \lambda^2 \left| H_1^0 \right|^2 \left| H_2^0 \right|^2 + \left| H_2^0 \right|^2 \sum_i \left( Y_{\nu}^i \right)^2 \left| \tilde{\nu}_i^c \right|^2  + \lambda^2 \left( \left| H_1^0 \right|^2 + \left| H_2^0 \right|^2 \right)  \left| \sum_i  \snuc_i \right|^2 + \kappa^2 \sum_i \left| \tilde{\nu}_i^c \right|^4 . \nonumber
\end{align}
We exchange the three parameters $\left\{ m_{H_1}^2, m_{H_2}^2, m_{\snuc}^2 \right\}$ for the real VEVs $\left\{ v_1 (0), v_2 (0), v_{\snuc} (0) \right\}$ by solving the three minimization equations 
\begin{align}\label{eq:MinEqns}
	\left. \frac{\partial V_0}{\partial H_1^0} \right|_{\mathrm{VEV}} &= 0 = m_{H_1}^2 v_1  + \frac{g_1^2+g_2^2}{4} \left( v_1^2 - v_2^2 \right) v_1 + 3 \left[ -a_{\lambda} v_2^{} v_{\snuc}^{} - \kappa \lambda v_2^{} v_{\snuc}^2 \right] + 3 \lambda^2 v_1 v_2^2 + 9 \lambda^2 v_1 v_{\snuc}^2 \nonumber \\
	\left. \frac{\partial V_0}{\partial H_2^0} \right|_{\mathrm{VEV}} &= 0 = m_{H_2}^2 v_2  - \frac{g_1^2+g_2^2}{4} \left( v_1^2 - v_2^2 \right) v_2 + 3 \left[ -a_{\lambda} v_1^{} v_{\snuc}^{} - \kappa \lambda v_1^{} v_{\snuc}^2 \right] + 3 \lambda^2 v_2 v_1^2 + 9 \lambda^2 v_2 v_{\snuc}^2 \nonumber \\
	\left. \frac{\partial V_0}{\partial \snuc_i} \right|_{\mathrm{VEV}} &= 0 = m_{\snuc}^2 v_{\snuc} +  \left[ -a_{\lambda} v_1^{} v_2^{} + a_{\kappa} \left( v_{\snuc}^{2} \right)^{} - 2  \kappa \lambda v_1 v_2 v_{\snuc}^{}  \right] + 3 \lambda^2 \left(  v_1^2 + v_2^2 \right) v_{\snuc} + 2 \kappa^2 v_{\snuc} v_{\snuc} ^2  
\end{align}
where ``VEV'' represents evaluating the fields at the zero temperature vacuum $\left\{ H_1^0, H_2^0, \snuc_i \right\} = \left\{ v_1 (0), v_2 (0), v_{\snuc} (0) \right\}$.  Terms proportional to $Y_{\nu}^i$ and $v_{\snu}$ are negligible and have been omitted.  Because of the $\mathbb{S}_3$ permutation symmetry of our potential, the three equations associated with the sneutrino field directions are identical.  

The one-loop radiative correction to the effective potential is given by the Coleman-Weinberg potential \cite{Coleman:1973jx} as a function of the field-dependent mass matrices $M^2_i$ calculated in the Landau gauge ($\xi = 0$).  The mass matrices which we use are included in Appendix \ref{sec:massmatrices} along with $n_i$, the degrees of freedom associated with each matrix that correspond to suppressed indices (negative for fermions).  Regulating UV divergences in $d = 4 - 2 \epsilon$ dimensions, the Coleman-Weinberg potential becomes
\begin{align}\label{eq:ColemanWeinberg}
	\Delta V_1^0 = \frac{1}{64 \pi^2} \sum_i n_i \mathrm{Tr}~M_i^4 \left(  \log \frac{M_i^2}{\mu^2} - \frac{3}{2} -C_{\mathrm{UV}} \right)
\end{align}
where $C_{\mathrm{UV}} = \frac{1}{\epsilon} - \gamma_E + \ln 4 \pi$ and $\mu$ is the t'Hooft scale.  We impose a mixed renormalization scheme in which the counterterms for the parameters $\left\{ m_{H_1}^2, m_{H_2}^2, m_{\snuc}^2 \right\}$ are chosen such that the zero temperature vacuum is unshifted by the radiative corrections.  This condition is equivalent to requiring tadpole graphs to vanish and imposes
\begin{align}\label{eq:SoftMassShifts}
	\delta  m_{H_1}^2 & =- \frac{1}{v_1} \left. \frac{\partial \Delta V_1^0}{\partial \left( H_1^0 \right)^{\ast}} \right|_{\mathrm{VEV}} \nonumber \\
	\delta  m_{H_2}^2 & =- \frac{1}{v_2} \left. \frac{\partial \Delta V_1^0}{\partial \left( H_2^0 \right)^{\ast}} \right|_{\mathrm{VEV}} \nonumber \\
	\delta  m_{\snuc}^2 & =- \frac{1}{v_{\snuc}} \left. \frac{\partial \Delta V_1^0}{\partial \left( \snuc_i \right)^{\ast}} \right|_{\mathrm{VEV}}  .
\end{align}
The remaining parameters are determined by the $\overline{DR}$ scheme and all parameters are specified at a renormalization scale of $\mu = 100$ GeV.  We make no assumptions about dominant contributions to the one-loop corrections but instead calculate \eref{eq:ColemanWeinberg} by summing all species that couple to the Higgs sector.  

At finite temperature, the effective potential receives an additional one-loop correction 
\begin{align}\label{eq:DV1T}
	\Delta V_1^T = \frac{T^4}{2 \pi^2} \left[ \sum_b n_b \mathrm{Tr}~J_{B} \left( M_b^2 / T^2 \right) + \sum_f n_f \mathrm{Tr}~J_{F} \left( M_f^2 / T^2 \right)  \right]
\end{align}
where the sums run over bosonic ($b$) and fermionic ($f$) mass matrices.  The thermal functions can be expressed as a a sum of modified Bessel functions of the second kind,
\begin{align}\label{eq:ThermalFuncs}
	J_B \left( y \right) & = \int_0^{\infty} dx \,  x^2 \log \left( 1 - e ^{- \sqrt{x^2 + y}} \right)  = -\sum_{n=1}^{\infty} \frac{1}{n^2} y K_2 \left( n \sqrt{y} \right) \\
	J_F \left( y \right) & = \int_0^{\infty} dx \,  x^2 \log \left( 1 + e ^{- \sqrt{x^2 + y}} \right) = -\sum_{n=1}^{\infty} \frac{\left( -1 \right)^n}{n^2} y K_2 \left( n \sqrt{y} \right)  . \nonumber
\end{align}
Because these integrals are computationally taxing, we use the Bessel function representation and truncate the sum at five terms.  This is a very good approximation and introduces less than 1\% of error.  

At high temperatures, the perturbative expansion fails unless higher order ``daisy'' graphs which diverge quadratically with temperature are resummed.  This procedure effectively replaces the bosonic field-dependent mass matrix $M_b^2$ with $M_b^2 + \Pi_b$ where $\Pi_b \propto T^2$ and amounts to including a term into the potential given by
\begin{align}\label{eq:Vdaisy}
	\Delta V_{\mathrm{daisy}} = - \frac{T}{12 \pi} \sum_b n_b \mathrm{Tr} \left[ \left( M_b^2 + \Pi_b \right)^{3/2} - \left( M_b^2 \right)^{3/2} \right] .
\end{align}
The thermal mass corrections $\Pi_b$ are included in Appendix
\ref{sec:bosonicthermalmasses}.  Combining all of the radiative and
finite temperature terms, the one-loop finite temperature effective
potential plus daisy resummation is given by
\begin{align}\label{eq:VeffT}
	V_{\mathrm{eff}}^T \left( T \right) & = V_0 + \Delta V_1^0 + \Delta V_1^T \left(  T \right) + \Delta V_{\mathrm{daisy}}  \left( T \right) .
\end{align}

The main difference between the NMSSM and the $\mu\nu$SSM relevant for strongly
first order EWPT can be summarized as follows:
\begin{enumerate}
\item Because of the multidimensionality of the singlet field space
  $\{v_{\tilde{\nu}_{i}^{c}}(T)\}$, there can be electroweak phase
  transitions accompanied by rotations within the singlet field
  space. This opens up a new class of phase transitions that are
  unlike any of the NMSSM transitions.  For example, the phase
  transition can take place with the singlet VEV hopping from one
  nonzero value to another: \begin{equation}
    \{v_{\tilde{\nu}_{i}^{c}}(T_{c})=x_{i}\neq0,\langle
    H_{j}\rangle=0\}\longrightarrow\{v_{\tilde{\nu}_{i}^{c}}(T_{c})=y_{i}\nparallel
    x_{i},\langle
    H_{j}\rangle\neq0\}.\label{eq:VEVhopping}\end{equation} The terms
  in Eqs.~(\ref{eq:superpotential}) and (\ref{eq:softterms}) that will
  play a particularly important role for this rotational hopping are
  the soft terms
  $-a_{\lambda}H_{1}\cdot H_{2}\tilde{\nu}_{i}^{c}+\frac{1}{3}a_{\kappa}\left(\tilde{\nu}_{i}^{c}\right)^{3}$
  (which control the cubic and lower dimension tree level couplings
  which in turn control radiative corrections) and the
  superpotential terms $-\lambda
  \hat{H}_{1} \cdot \hat{H}_{2} \hat{\nu}_{i}^{c}+\frac{1}{3}\kappa\left(\hat{\nu}_{i}^{c}\right)^{3}$
  (which control the quartic and lower dimension tree level
  couplings).\footnote{As we will see later, the shift of the field
    origin will generically generate lower dimension couplings from
    higher dimensional couplings.}

\item There is a soft term coupling the singlet to the up-type Higgs,
\begin{equation}
\Delta V_{\mathrm{soft}}\sim a_{\nu} \snu_i H_2^0 \tilde{\nu}^{c}_i ,
\end{equation}
which potentially provides a cubic coupling for the Higgs sector.
Unfortunately, this term does not play an important role in the analysis
because $v_{\tilde{\nu}}\ll\mathcal{O}(\mbox{GeV})$.

\item The superpotential has a Yukawa coupling of the singlet to the left-handed
lepton and Higgs, leading to the following additional $F$-terms:
\begin{eqnarray}\label{eq:smallyukawaeffects}
 	\Delta V_{F} & = & Y_{\nu} \sum_{i} \left[ 
	\kappa  \left( H_2^0 \snu_i \right)^{\ast} \left( \snuc_i \right)^2
	- \lambda  H_1^0 \left( 
	\left| H_2^0 \right|^2  \snu_i^{\ast}
	+   \sum_j \left( \snu_j \snuc_j \right)^{\ast}  \snuc_i	
	\right)
	+ \hc \right]\nonumber \\
 &  & + Y_{\nu}^{2} \left( \left| H_2^0 \right|^2 \sum_i \left( \left| \snu_i \right|^2 +  \left| \snuc_i \right|^2 \right) + \left| \sum_i \snu_i \snuc_i \right|^2 \right)
 \end{eqnarray}
Given the smallness of the Yukawa couplings $Y_{\nu}\sim\mathcal{O}(10^{-7})$,
these terms are not particularly important for the phase transition
when the transition occurs with VEVs of order $10^7$ GeV or less. Note that
all of these terms are quartic in nature owing to the absence of dimensionful
parameters in the superpotential. Also, since the origin of R-parity
violation is the leptonic Yukawa coupling which is also the source
of $\Delta V_{F}$ , we see that these do not play a significant role.
\end{enumerate}
Hence, a generic feature of the $\mu\nu$SSM SFOPT not reproducible by the
NMSSM is the feature due to point 1 above. This will be emphasized in
the numerical exploration below.  We will also find one step
transitions, which are qualitatively similar to the NMSSM
transitions.

\section{\label{sec:qualitativedesc}Qualitative Description of The Desired Parametric Region}
A novel feature of the $\mu\nu$SSM compared to the NMSSM is the transition
depicted in Eq.~(\ref{eq:VEVhopping}). In such cases one can shift the
origin of the field such that the phase transition of interest occurs
from the origin. With such shifted coordinates in mind, we define the
field $\phi$ to be the radial magnitude \begin{equation}
  \phi\equiv\sqrt{(\vec{v})^{2}+(\Delta\vec{v}_{\tilde{\nu}})^{2}+(\Delta\vec{v}_{\tilde{\nu}^{c}})^{2}}\label{eq:shiftedmagnitude}\end{equation}
for a phase transition controlled by the potential $V_{T}(\phi)$ in
which the vector of CP-even Higgs scalars attains an order parameter
change of $\vec{v}$. Explicitly, the strength of the phase transition
is approximately characterized by the $SU(2)_{L}$ breaking
$|\vec{v}|/T_{c}$ and not $\phi_{c}/T_{c}$ where the critical
temperature $T_{c}$ is defined by the condition $V_{T_{c}}(0)\approx
V_{T_{c}}(\phi_{c})$.

\begin{figure}[ht]
\begin{centering}
\includegraphics{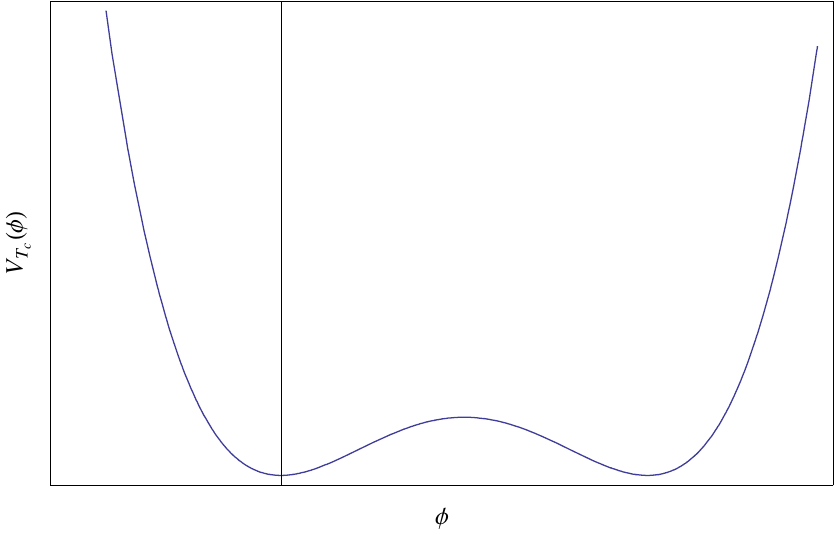}
\par\end{centering}
\caption{\label{fig:A-schematic-plot}A schematic plot of the finite temperature
effective potential at the critical temperature of Eq. (\ref{eq:criticaltemp1}).
The vertical line represents $\phi=0$ and helps to visualize the
effect of $\phi\rightarrow-\phi$ symmetry breaking effect of the
cubic term which is responsible for the bump at $\phi>0$ for $\{E>0,F_{\mathrm{na}}=0\}$
in Eq. (\ref{eq:finite T effective potential raw}).}
\end{figure}

The finite temperature corrected effective potential of a real scalar
field $\phi$ near the critical temperature will behave approximately
as\begin{equation}
V_{T}(\phi)\approx(\frac{1}{2}M^{2}+c_{1}T^{2})\phi^{2}-E_{\eff}\phi^{3}+F_{\mathrm{na}}(\phi,T)+\frac{\lambda_{\eff}}{4}\phi^{4}+\mbox{other temperature dependences}\label{eq:finite T effective potential raw}\end{equation}
where $c_{1}\sim\mathcal{O}(1)$ constant proportional to coupling
constants responsible for the leading mass correction and $F_{\mathrm{na}}$
is the nonanalytic thermal correction contribution that can lead
to an effective cubic contribution to the potential. Although in the 
MSSM $F_{\mathrm{na}}$ plays a significant role, with a singlet involved
such as in the $\mu\nu$SSM, $F_{\mathrm{na}}$ need not play a crucial
role. Hence, we will set $F_{\mathrm{na}}=0.$ In this section, we neglect
{}``other temperature dependences'' in Eq.~(\ref{eq:finite T effective potential raw}).
Note that Eq.~(\ref{eq:finite T effective potential raw}) has $M^{2}>0$
even though at $T=0,$ symmetry is broken when $E_{\eff}>0$
and satisfies a condition specified below. 

Defining where $v(T_{c})$ is the degenerate minimum VEV, we find
\begin{eqnarray}
T_{c} & = &
\frac{\sqrt{2E_{\eff}^{2}/\lambda_{\eff}-M^{2}}}{\sqrt{2c_{1}}}.\label{eq:criticaltemp1}\end{eqnarray}
The potential at the critical temperature is depicted in
Fig. \ref{fig:A-schematic-plot}.  The unusual sign of
$2E_{\eff}^{2}/\lambda_{\eff}-M^{2}$ stems from our
assumption that the symmetry is broken at zero temperature due to
predominantly the $E_{\eff}$ term. This situation turns out to
be generically beneficial for a SFOPT as we explain soon below.  The
critical temperature $T_{c}$ is larger if $E_{\eff}>0$ because
in that case, the negative contribution from the cubic term in
Eq. (\ref{eq:finite T effective potential raw}) is enhanced for
$\phi>0$ which means that the quadratic term which is the leading
source of positivity (as $\phi$ approaches $\phi_{c}$ from the left)
has to be stronger to cancel the stronger negative contribution. Since
there will be no positive mass squared at the origin during the phase
transition in the absence of the cubic term, the mass at the origin
has to be also larger for increasing $E_{\eff}>0$.  Explicitly, the
mass at the origin (which by construction is our starting point of the
phase transition) is\begin{equation}
\partial_{\phi}^{2}V_{T=T_{c}}(0)=\frac{2E_{\eff}^{2}}{\lambda_{\eff}}.\label{eq:masstermatorigin}\end{equation}
This mass is identical to the mass at $\phi=\phi_{c}$.
We can also understand the VEV\begin{eqnarray}
\phi_{c} & = & \frac{2E_{\eff}}{\lambda_{\eff}}\end{eqnarray}
which can be heuristically justified by the fact that the broken phase
local minimum results from a competition between the cubic and the
quartic term (which is the dominant source of positivity as $\phi\rightarrow\phi_{c}^{+}$)
at the time of critical temperature when the mass term is again controlled
by Eq.~(\ref{eq:masstermatorigin}). 

Finally, the strength of the $SU(2)_{L}$ breaking in the transition
is given by \begin{eqnarray}
\frac{v(T_{c})}{T_{c}} & = & \frac{v(T_{c})\sqrt{2c_{1}}}{M\sqrt{\frac{2E_{\eff}^{2}}{\lambda_{\eff}M^{2}}-1}}\label{eq:exact}\end{eqnarray}
where 
\begin{equation}
v(T_{c})=\phi_{c}f(\vec{\Omega})\end{equation}
and $f(\vec{\Omega})$ is a projection cosine onto the Higgs axis.
By definition of the $SU(2)_{L}$ breaking transition, $\phi_{c}f(\vec{\Omega})\lesssim\mathcal{O}(v(0))$.
At $T=0$, $\phi(0)\equiv\langle\phi\rangle_{T=0}$ is related to
$M$ through \begin{equation}
M=\sqrt{3E_{\eff}\phi(0)-\lambda_{\eff}\phi^{2}(0)}\end{equation}
where $\lambda_{\eff}\phi(0)>3E_{\eff}$. Equation (\ref{eq:exact})
thus can be rewritten in terms of $\phi(0)$ as \begin{equation}
\frac{v(T_{c})}{T_{c}}=\left(\frac{2E_{\eff}}{\lambda_{\eff}\phi(0)}\right)\frac{\sqrt{2c_{1}/\lambda_{\eff}}}{\sqrt{\left(1-\frac{E_{\eff}}{\lambda_{\eff}\phi(0)}\right)\left(1-\frac{2E_{\eff}}{\lambda_{\eff}\phi(0)}\right)}}f(\vec{\Omega}).\label{eq:phasetransparamdep}\end{equation}
Hence, the strength of the phase transition is controlled mostly by
2 parameters: \begin{equation}
\{\frac{E_{\eff}}{\lambda_{\eff}\phi(0)},\sqrt{\frac{c_{1}}{\lambda_{\eff}}}\}.\end{equation}

Note that since $f(\vec{\Omega})\leq1$, this angular projection function
can only enhance the phase transition in a limited manner. Requiring
$\frac{v(T_{c})}{T_{c}}$ be real and $V(\phi(0))<V(0)$
result in the condition\begin{equation}
0\leq\frac{E_{\eff}}{\lambda_{\eff}\phi(0)}\leq\frac{1}{2}.\label{eq:Eoverlamv0bound}\end{equation}
Therefore, one should keep in mind that although having a nonvanishing
$E_{\eff}$ is good for a strong first order phase transition,
the enhancement is bounded. Indeed, this bound is approximately satisfied
by the numerical analysis, and SFOPT points that we find occur when
\begin{equation}
\frac{E_{\eff}}{\lambda_{\eff}\phi(0)}\approx\frac{1}{2}.\label{eq:crucialsaturate}\end{equation}
From the derivation of Eq.~(\ref{eq:phasetransparamdep}), one can
see that Eq.~(\ref{eq:crucialsaturate}) corresponds to making $T_{c}$
as small as possible during the phase transition. When $\frac{E_{\eff}}{\lambda_{\eff}\phi(0)}>\frac{1}{2}$
the origin becomes the global minimum and the symmetry is unbroken.
Note also that because $\phi$ is defined with respect to the shifted
singlet origin in Eq.~(\ref{eq:shiftedmagnitude}), $\phi(0)$ does not
correspond to the radial magnitude of the scalar field from the
original Lagrangian's field origin.

After this first order phase transition, a second order phase transition
might occur when $V''(0)=0.$ However, with $M^{2}>0$,
this does not occur for this 1D toy model. Note that \cite{Pietroni:1992in}
assumes that there exists a temperature for which $V''=0$ which in
fact never occurs for this toy model.

Generically, we are interested in a strong first order phase transition
characterized by 
\begin{equation}
\sqrt{2} \frac{v(T_{c})}{T_{c}}\gtrsim1.3\label{eq:strongfirstorder}\end{equation}
\cite{Bochkarev:1990gb}. If the asymptotic conditions $\frac{E_{\eff}}{\lambda_{\eff}\phi(0)}\rightarrow1/2$
and/or $c_{1}/\lambda_{\eff}\rightarrow\infty$ are met, the
phase transition can be arbitrarily strong. However, the following
phenomenological constraints prevent/constrain an arbitrarily strong
transition in Eq. (\ref{eq:phasetransparamdep}):
\begin{enumerate}
\item Global minima shifts can prevent the saturation of $E_{\eff}/\left[\lambda_{\eff}\phi(0)\right]=1/2$
for a particular underlying parametric path. For example, as one approaches
$E_{\eff}/\left[\lambda_{\eff}\phi(0)\right]=1/2$ within
a particular region of underlying parameter space,
\footnote{Recall that $E_{\eff}/\left[\lambda_{\eff}\phi(0)\right]$
are effective parameters derivable from underlying Lagrangian parametrized
for example as Eqs.~(\ref{eq:superpotential}) and (\ref{eq:softterms}).
} the origin of Eq.~(\ref{eq:shiftedmagnitude}) has to be shifted
to a new global minimum (where the electroweak symmetry is still not
broken, i.e. $\vec{v}=0$). When this occurs, $\{E_{\eff},\lambda_{\eff},c_{1}\}$
of Eq.~(\ref{eq:finite T effective potential raw}) undergo a discontinuous
change as a function of the underlying parameters such as those of
Eqs.~(\ref{eq:superpotential}) and (\ref{eq:softterms}).
\item Small $\lambda_{\eff}$ can result in phenomenologically unacceptably
light Higgs (or other scalar masses). For example, it is clear from
the effective model that \begin{equation}
m_{\phi}^{2}=2\lambda_{\eff}\phi^{2}(0)\left[1-\frac{3}{2}\frac{E_{\eff}}{\lambda_{\eff}\phi(0)}\right]\label{eq:Higgs mass}\end{equation}
where the term in the parenthesis in Eq. (\ref{eq:Higgs mass}) is
positive since $0\leq E_{\eff}/\left(\lambda_{\eff}\phi(0)\right)\leq1/2$.
Note that in any models that embed the MSSM, there is a minimal contribution
to $\lambda_{\eff}$ from the D-terms that also makes it difficult
to make it arbitrarily small. Note also that increasing $\frac{E_{\eff}}{\lambda_{\eff}\phi(0)}$
lowers the $\phi$ mass as well.
\item When $E_{\eff}/(\lambda_{\eff}\phi(0))\rightarrow1/2$,
the energy difference $\Delta V$ between the false vacuum and true
vacuum asymptotically vanishes. Explicitly, we have as $E_{\eff}\rightarrow E_{c}\equiv\lambda_{\eff}\phi(0)/2$,
we find \begin{equation}
\Delta V\rightarrow\frac{2\sqrt{2}\Delta E_{\eff}M^{3}}{\lambda^{3/2}}\rightarrow0\end{equation}
\begin{equation}
\frac{v(T_{c})}{T_{c}}\rightarrow\left(\frac{2}{\lambda_{\eff}}\right)^{1/4}\sqrt{\frac{c_{1}M}{\Delta E_{\eff}}}\end{equation}
where $\Delta E_{\eff}\equiv E_{\eff}-E_{c}$. Since the
validity of this estimate requires $\Delta V>0$, this region of parameter
space becomes very sensitive to radiative corrections. 
\item The contributions to $c_{1}$ \emph{that maximize}
  $\sqrt{c_{1}/\lambda_{\eff}}$ typically contribute to
  $\lambda_{\eff}$ as well (with different powers). Hence,
  particularly in the $\mu\nu$SSM, we are in a region where
  $\lambda_{\eff}$ is on the larger side and not the small
  side. 

\end{enumerate}
The features just discussed qualitatively explain the numerical scan
of the parameter space which identifies a particular parametric region
in which Eq. (\ref{eq:strongfirstorder}) is satisfied at the same time
with some basic phenomenological constraints which we detail in
Sec.~\ref{sec:paramscan}.  There, more analytic formulas will be given
explaining some of the features of the numerical results.

Now, let us consider the general path of the electroweak phase
transition.  At $T\gg\mathcal{O}(\mbox{TeV})$, the global minimum will
be at \begin{equation}
  \{v_{\tilde{\nu}_{i}^{c}}(T)=0,v_{i}(T)=0\},\end{equation} the
scalar field origin.\footnote{Symmetry restoration typically occurs as
  long as there are no tadpole contribution proportional to the
  temperature.}  As explained previously, the left-handed slepton VEVs
are undergoing small energy scale transitions which are not
particularly relevant to most of the discussion. As the temperature is
lowered, a nontrivial singlet VEV configuration will realize a global
minimum, and the system will consequently make a transition. This
transition in the singlets is sometimes accompanied by an electroweak
symmetry breaking phase transition and sometimes not. If the first
nontrivial singlet transition is accompanied by strongly first order
electroweak symmetry breaking, these would be SFOPT from the scalar
field origin:\begin{equation}
  \{v_{\tilde{\nu}_{i}^{c}}(T_{c}^{+})=0,v_{j}(T_{c}^{+})=0\}\longrightarrow\{v_{\tilde{\nu}_{i}^{c}}(T_{c}^{-})\neq0,v_{j}(T_{c}^{-})\neq0\}.\end{equation}
In this case, the origin of the vector whose magnitude is taken in
Eq.~(\ref{eq:shiftedmagnitude}) will be zero.

In addition, there will generically be singlet transitions from the
origin at temperature $T_{O}$ first without an electroweak phase
transition, of the form\begin{equation}
\{v_{\tilde{\nu}_{i}^{c}}(T_{O}^{+})=0,v_{j}(T_{O}^{+})=0\}\longrightarrow\{v_{\tilde{\nu}_{i}^{c}}(T_{O}^{-})=x_i(T_{O})\neq0,v_{j}(T_{O}^{-})=0\}.
\label{eq:originshifttrans}
\end{equation}
Even if this is a first order phase transition, it will typically
complete before the subsequent electroweak symmetry breaking, and thus
it does not, to leading approximation, participate in EWBG.  However, it
can in principle be relevant for gravity waves (see
e.g. \cite{Steinhardt:1981ct,Witten:1984rs,Kosowsky:1991ua,Kosowsky:1992rz,Kamionkowski:1993fg,Apreda:2001us,Kosowsky:2001xp,Dolgov:2002ra,Nicolis:2003tg,Caprini:2006jb,Grojean:2006bp,Randall:2006py,Gogoberidze:2007an,Huber:2007vva,Caprini:2007xq,Megevand:2008mg,Espinosa:2008kw,Huber:2008hg,Caprini:2009fx,Caprini:2009yp,Kusenko:2009cv,Megevand:2009ut,Ashoorioon:2009nf,Das:2009ue,Kahniashvili:2009mf,Kehayias:2009tn,Durrer:2010xc,Chung:2010cb}).
Afterwards, there is a subsequent electroweak symmetry breaking phase
transition
\begin{equation}
\{v_{\tilde{\nu}_{i}^{c}}(T_{c}^{+})=x_i(T_c)
\neq0,v_{j}(T_{c}^{+})=0\}\longrightarrow\{v_{\tilde{\nu}_{i}^{c}}(T_{c}^{-})=y_i\neq
0,v_{j}(T_{c}^{-})\neq0\}
\label{eq:secondsteptrans}
\end{equation}
whose strength is important for EWBG. In this case, the origin of
the vector whose magnitude is taken in Eq.~(\ref{eq:shiftedmagnitude})
will be
$\{v_{\tilde{\nu}_{i}^{c}}(T_{c}^{+})=x_i,v_{j}(T_{c}^{+})=0\}$.  
When $x_i \nparallel y_i$, this transition corresponds to a
``rotation'' of the singlet vector.

Before concluding this section, let us briefly describe how the
discrete symmetry discussed below Eq.~(\ref{eq:softterms}) and zero
temperature radiative corrections plays a role for some of our strong
multistep transitions.  Once a phase transition of the form
Eq.~(\ref{eq:originshifttrans}) takes place, the set of degenerate
global minima will form a coset representation of $\mathbb{Z}_3
\otimes \mathrm{CP} \otimes \mathbb{S}_3$.\footnote{Recall that $\mathbb{S}_3$ is nearly an exact
  symmetry because of the smallness of the leptonic Yukawa coupling.}
Because of the approximate $\mathbb{Z}_3 \otimes \mathbb{Z}_3 \otimes
\mathbb{Z}_3$ symmetry described in Eq.~(\ref{eq:Z3independent}), to
tree level accuracy, the coset space will be actually bigger:
$\mathbb{Z}_3 \otimes \mathbb{Z}_3 \otimes \mathbb{Z}_3 \otimes \mathrm{CP}
\otimes \mathbb{S}_3$.  Some of the $\mathbb{Z}_3 \otimes \mathbb{Z}_3 \otimes
\mathbb{Z}_3$ minima will be split due to the zero temperature
radiative corrections, and the global minimum will be at a subset of the
$\mathbb{Z}_3 \otimes \mathbb{Z}_3 \otimes \mathbb{Z}_3$ minima [one
of which is what we labeled as $\vec{x}(T_O)$ in
Eq.~(\ref{eq:originshifttrans})].  Finally, when the temperature drops
enough to make one of the EWSB minima degenerate with $\vec{x}(T_c)$, the
transition depicted by Eq.~(\ref{eq:secondsteptrans}) occurs.

In Sec.~\ref{sec:paramscan}, we will discuss explicit examples of both
one step and multistep phase transitions.

\section{\label{sec:sphaleron} Weak Sphaleron and the Singlet}

After the baryon asymmetry has been created at a first order
electroweak phase transition, it may be washed out by the B-violating
sphaleron process \cite{Klinkhamer:1984di, Kuzmin:1985mm} in the
broken phase.  The sphaleron is a nonperturbative field configuration
in the Weinberg-Salam theory that interpolates between topologically
distinct vacua and violates $B+L$.  To avoid washout, one must require
that sphaleron transitions are suppressed meaning that the rate of
these processes is less than the Hubble parameter at the time of the
phase transition.  This imposes a lower bound on the sphaleron
Euclidean action $E_{\sph} (T_c)/ T_c \gtrsim 45$
\cite{Bochkarev:1990gb} which, in the Standard Model, becomes a lower
bound on the Higgs VEV in the broken phase $\sqrt{2} v (T_c) / T_c
\gtrsim 1.3$ where $v(0) = 174$ GeV.  The six sneutrino fields of the
$\mu \nu$SSM which receive VEVs during EWSB could in principle modify
this bound.  As we will see, the modifications are small because 1)
the left-handed sneutrino VEV is much less than the electroweak scale
and 2) the singlet sneutrino has a nearly homogenous solution which
stays nears the minimum of the potential.\footnote{This second conclusion was pointed out by Ref.~\cite{Ahriche:2007jp} in which the author calculated the weak sphaleron in the SM extended by a real singlet.}  

To obtain the sphaleron action at finite temperature, we calculate the zero temperature sphaleron and apply the scaling law \cite{Brihaye:1993ud}
\begin{align}\label{eq:ScalingLaw}
	E_{\sph} (T) = E_{\sph} (0) \frac{v(T)}{v(0)} 
\end{align}
which introduces less than a 10\% error.  
Additionally, we compute the sphaleron solution using the tree level scalar potential $V_0$ and neglect radiative corrections.  
To a very good approximation \cite{Klinkhamer:1984di} we can also neglect the $\U{1}_Y$ gauge coupling and compute a purely $\SU{2}_L$ sphaleron solution.  The sphaleron ansatz is static and possesses an $\SO{3}$ rotational symmetry.  The ansatz is given by 
\begin{align}\label{eq:SphaleronAnsatz}
	\left\{ H_1,H_2, \tilde{L}_i \right\} & = \left\{ v_1 h_1(\xi), v_2 h_2(\xi), v_3 h_3(\xi) \right\}  U^{\infty} \left( \begin{array}{c}0\\1\end{array} \right) \nonumber \\
	\snuc_i &= v_4 h_4(\xi) \nonumber \\
	W_i^a \sigma^a dx^i & = -\frac{2i}{g} f \left( \xi \right) dU^{\infty} (U^{\infty})^{-1}  \\
	U^{\infty} & = \frac{1}{r} \left( \begin{array}{c c}z&x+iy\\-x+iy&z\end{array} \right)
\end{align}
in terms of the dimensionless radial coordinate $\xi = r/r_0$ rescaled
by $r_0 = \left( g \sqrt{2} \sqrt{v_1^2 + v_2^2 + v_{\snu}^2}
\right)^{-1} \approx \left( g \sqrt{2} v \right)^{-1}$.  All VEVs are
evaluated at zero temperature and we have introduced $v_3 = v_{\snu}$
and $v_4 = v_{\snuc}$ for convenience.  We have used the
$\mathbb{S}_3$ symmetry to equate the functions that describe the
sneutrino fields of different generations, such that the sphaleron
solution is given by five functions $h_i \left( \xi \right)$ and $f
\left( \xi \right)$.  With this ansatz the field equations become
\begin{align}\label{eq:SphaleronEOM}
\frac{\xi^2 r_0^2}{v_i^2} \frac{\partial V_0}{\partial h_i} = 
\begin{cases}
	\xi^2 h_i^{\prime \prime} + 2\xi h_i^{\prime} - 2(1-f)^2 h_i & i = 1, 2 \\
	3 \left[ \xi^2 h_3^{\prime \prime} + 2\xi h_3^{\prime} - 2(1-f)^2 h_3 \right] & i = 3 \\
	3 \left[ \xi^2 h_4^{\prime \prime} + 2\xi h_4^{\prime} \right] & i = 4
\end{cases} \nonumber \\
	\xi^2 f^{\prime \prime} - 2f(1-f)(1-2f) = -\frac{1}{4}\xi^2(1-f)\left(v_1^2 h_1^2 + v_2^2 h_2^2 +3 v_{3}^2 h_3^2 \right) g^2 r_0^2 .
\end{align}
where the potential is normalized to vanish in the EWSB vacuum $V_0 \left( h_i = 1 \right) = 0$.  Note that the term $(1-f)^2 h_i$ is absent for $i=4$ because the singlet sneutrinos do not couple to the gauge bosons.  The sphaleron action is obtained by integrating the sphaleron solution
\begin{align}\label{eq:SphaleronAction}
	E_{\sph} \left( 0 \right) =  \frac{ 4 \pi v \sqrt{2}}{g} \int_{0}^{\infty} & \left\{
	 4 \left( \frac{df}{d \xi} \right)^2 
	+ \frac{8}{\xi^2} f^2 \left( 1 - f \right)^2 
	+\xi^2  \frac{V_0 \left( h_i \right)}{g^2 v^4} \right. \nonumber \\
	 & \left. +\frac{\xi^2}{2 v^2} \left[  \left( v_1 \frac{d h_1}{d \xi} \right)^2 +  \left( v_2 \frac{d h_2}{d \xi} \right)^2 + 3 \left( v_3 \frac{d h_3}{d \xi} \right)^2 + 3 \left( v_4 \frac{d h_4}{d \xi} \right)^2 \right] \right. \nonumber \\
	& \left. + \frac{1}{v^2} \left( v_1^2 h_1^2 + v_2^2 h_2^2 + 3 v_3^2 h_3^2 \right) \left( 1-f \right)^2  
	\right\} d \xi
\end{align}
We can observe immediately that contributions from the left-handed sneutrinos will be negligible because the function $h_3$ always appears with a prefactor of $v_3  = v_{\snu}  \ll v$.  

We can study the sphaleron solution by considering the asymptotic limits of \eref{eq:SphaleronEOM}.  In the large $\xi$ limit, all five field profiles must asymptote to unity in order for $E_{\sph}$ to be finite.  In the small $\xi$ limit, we find that the gauge boson and the three weakly charged scalar functions asymptote to zero, as in the Weinberg-Salam model, but that the singlet function approaches a value which can in general be nonzero:
\begin{align}\label{eq:SphaleronSmallXi}
	f \left( \xi \right) \xrightarrow{\xi \to 0} \alpha \xi^2, 
	&&  h_i \left( \xi \right) \xrightarrow{\xi \to 0} \beta_i \xi  \ \ \ \ \   i \in \left\{ 1, 2, 3 \right\} ,
&& h_4  \xrightarrow{\xi \to 0} c + \beta_4 \xi^2	.
\end{align}
The singlet function behaves differently in this limit because the
gauge coupling term $(1-f)^2 h_4$ in the field equation is absent.
The boundary condition on the singlets makes the solution for $h_4
\left( \xi \right)$ qualitatively different than for the Higgs fields.
In particular, the solution $h_4 \left( \xi \right)$ which minimizes
$E_{\sph}$ will tend to be homogenous with $h_4 \approx 1$ for all
$\xi$.  The solution is homogenous because $E_{\sph} \ni \left( d h_4
/ d \xi \right)^2$ is positive semidefinite.  Hence, it can be
minimized by a constant $h_4$, and the solution remains near $h_4 = 1$
because this is where $V_0$ is minimized.  As a result, the singlet
fields contribute negligibly to the sphaleron action.

\begin{figure}[h]
\begin{center}
\includegraphics[width=0.45 \textwidth]{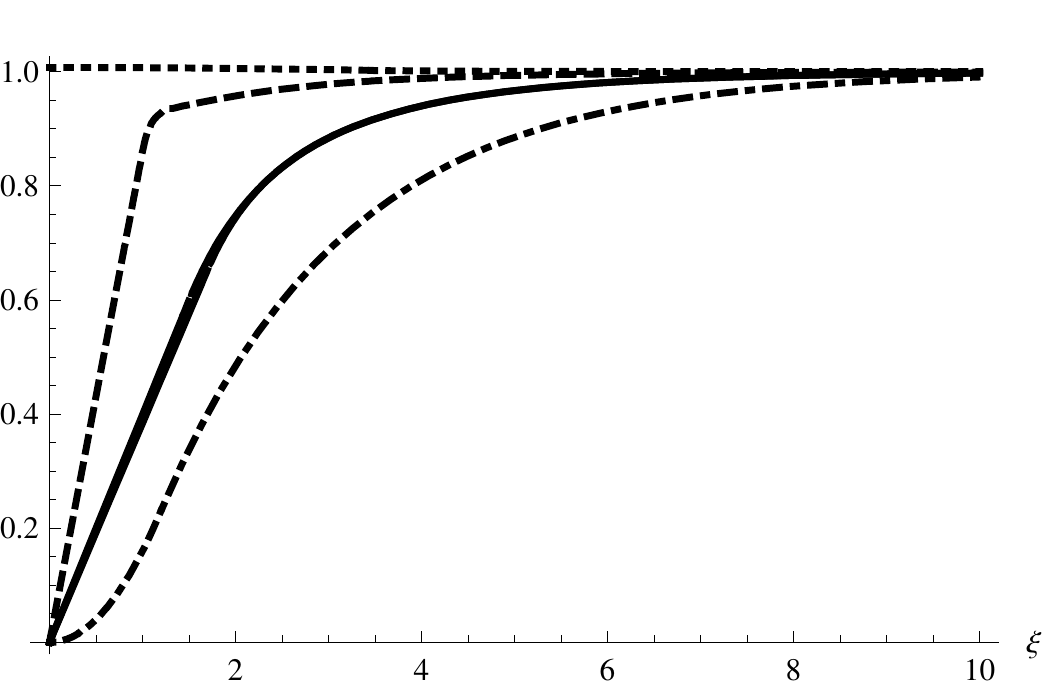} \hfill
\includegraphics[width=0.45 \textwidth]{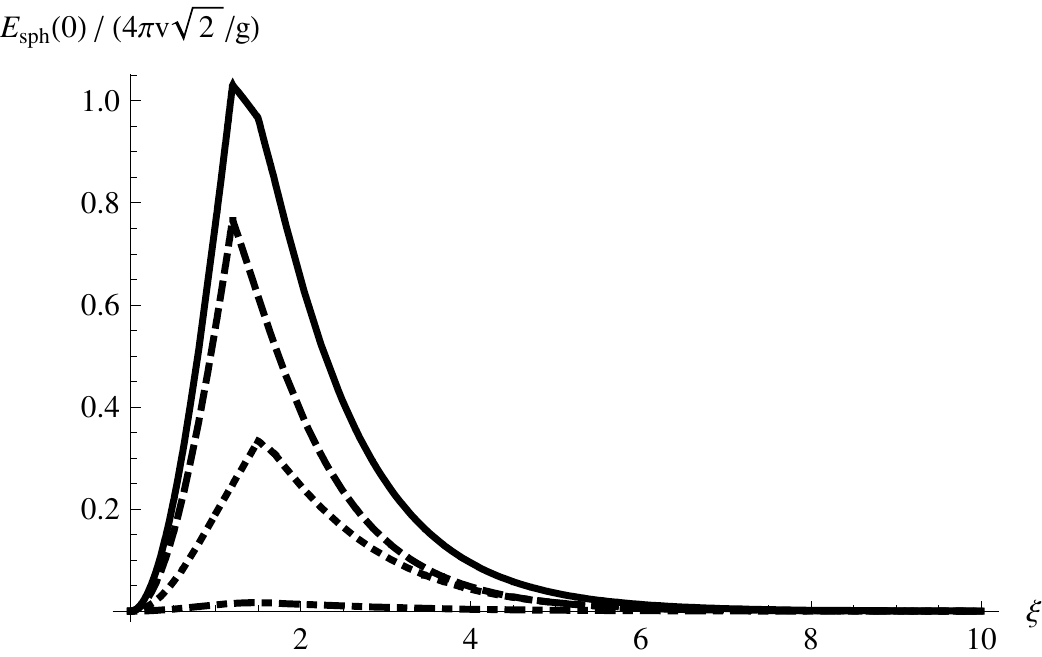}
\caption{\label{fig:Sphaleron} On the left, the $\mu \nu$SSM sphaleron
  solution versus the dimensionless radial coordinate with $h_1$ and
  $h_2$ (solid line), $h_3$ (dashed line), $h_4$ (dotted line), and $f$ (dashed-dotted line).  The
  solution $h_{\snuc}$ for the singlet sneutrinos does not satisfy the
  same boundary condition at $\xi \to 0$ as the $\SU{2}_L$ charged
  scalars. Hence, the solution of minimum energy is the one in which
  $h_{\snuc} \approx 1$ for all $\xi$.  On the right, the sphaleron
  energy density, \eref{eq:SphaleronAction}, with gauge kinetic terms
  (dashed line), scalar kinetic terms (dotted line), scalar potential terms
  (dashed-dotted line), and the total energy density (solid line).  This plot
  illustrates that the sphaleron action is dominated by the kinetic
  terms and that the contribution from the scalar potential is
  negligible.  }
\end{center}
\end{figure}

The sphaleron solution and energy density for a fiducial parameter set
are plotted in Fig. \ref{fig:Sphaleron}.  To obtain the field
profiles we solve \eref{eq:SphaleronEOM} in the large and small $\xi$
limits analytically, then match the solutions at five radii $r_i$
which are chosen to minimize $E_{\sph}$.  As discussed above, the
singlet solution hovers around $h_4 = 1$ where the potential has a
minimum.  To display how each term in \eref{eq:SphaleronAction}
contributes to the sphaleron action, we have also plotted the
integrand for the gauge kinetic, scalar kinetic, and scalar potential
contributions separately.  We observe that the sphaleron action is
dominated by the kinetic terms.  Since the parametric dependence only
appears explicitly in the scalar potential, which is negligible, we
expect that the sphaleron action is largely independent of our
parameter choice.  For this parameter set we find $E_{\sph} ( 0 )
\approx 1.83 \frac{4 \pi v}{g} \approx 8.7 \TeV$ which translates into
a bound on the Higgs VEV at the critical temperature that is $\sqrt{2}
\frac{v ( T_c )}{T_c} \gtrsim 1.3$.  As such, the Higgs VEV must
satisfy the same constraint in the $\mu \nu$SSM as in the SM to avoid
washout.

\section{\label{sec:paramscan}Parameter Scan and Phenomenological Bounds}

We have investigated the $\mu \nu$SSM phase transition by performing a two-dimensional parameter space scan.  For the two free parameters we use $m_{\ch} = 3 \lambda v_{\snuc}$, which coincides with the charged Higgsino mass in the limit $M_2 \gg m_W, m_{\ch}$, and the dimensionless variable
\begin{align}
	\sigma = 2 \left( \frac{a_{\lambda}}{m_{\ch} \lambda} + \frac{\kappa}{3 \lambda} \right) .
\label{eq:sigmadef}
\end{align}
These parameters are scanned uniformly over the ranges $m_{\ch} : \left[ 75
  \GeV, 175 \GeV \right]$ and $\sigma : \left[ 0, 25 \right]$ by
varying $v_{\snuc}$ and $a_{\lambda}$.  The SUSY-breaking parameters
are chosen to match the conventions of \cite{Escudero:2008jg}:
$\left\{ m_{\tilde{Q}}^2, m_{\tilde{u}^c}^2, m_{\tilde{d}^c}^2,
m_{\tilde{e}^c}^2 \right\}$ are fixed at a fiducial SUSY-breaking
scale which is taken to be 1 TeV, gaugino masses are set to $6 M_1 = 3
M_2 = M_3 = 3$ TeV, and A-terms are scaled by the associated Yukawa
couplings as $\left\{ A_{u_i} = a_u / Y_{u_i}, A_{d_i} = a_d/Y_{d_i},
A_{e_i} = a_e/Y_{e_i}, A_{\nu} = -a_{\nu} / Y_{\nu} \right\}$ and
fixed at 1 TeV.  Note that we have assumed for simplicity common
left-handed sneutrino VEVs $v_{\tilde{\nu_i}} = v_{\tilde{\nu}}$ and a
diagonal Yukawa matrix $\left( Y_{\nu} \right)_{ij} = Y_{\nu}
\delta_{ij}$.  The remaining soft masses $\left\{ m_{H_1}^2,
m_{H_2}^2, m_{\tilde{L}}^2, m_{\tilde{\nu}^c}^2 \right\}$ are
exchanged for the VEVs $\left\{ v_1 \left( 0 \right), v_2 \left( 0
\right), v_{\tilde{\nu}} \left( 0 \right), v_{\tilde{\nu}^c} \left( 0
\right) \right\}$ by solving the minimization equations at
zero temperature.  The remaining Higgs sector parameters are chosen to
be $\tan \beta = 2.6, \kappa = -0.64, \lambda = 0.18, v_{\tilde{\nu}}
= 1.4 \times 10^{-5}$ GeV, and $a_{\kappa} = -236$ GeV.  Given that
some of our sparticle masses are far larger than $T_c \sim
\mathcal{O}(100)$GeV, we could have integrated out these fields giving
rise to a more illuminating effective field theory parametrization
within the $\overline{DR}$ scheme.\footnote{ Recall that in
  $\overline{DR}$ scheme, decoupling is accomplished ``by hand''
  through computing threshold corrections after integrating out
  fields.}  However, to stay similar to the parametrization used in
\cite{LopezFogliani:2005yw,Escudero:2008jg}, and to give a relatively
unrestricted range for possible $T_c$, we have kept these relatively
heavy fields as dynamical.

At each point in the parameter space, we calculate the $\mu \nu$SSM
spectrum.  In order to get a handle on phenomenological constraints,
we impose the MSSM search bounds for the SUSY particles and require
the Higgs masses to be $\gtrsim 90$ GeV \cite{PDG} (later we will show
a sample parametric point Higgs spectrum with the lightest Higgs mass
of about $110$ GeV).
Model dependent bounds are of interest, but typically they are weaker,
as far as the neutral Higgs is concerned, because of singlet mixing
effects.  A more complete model dependent phenomenological consistency
check including the study of charged Higgs mediated $b \rightarrow s
\gamma$ rates is beyond the scope of this paper.  We calculate the
spectrum of the charged Higgses ($\phi_i^{\pm}$), charginos
($\tilde{\chi}_{i}^{\pm}$), and neutralinos ($\tilde{\chi}_i^0$) at
tree level and require $\left\{ m_{\phi_1^{\pm}} > 79.3 \mbox{ GeV},
m_{\tilde{\chi}_1^{\pm}} > 94 \mbox{ GeV}, m_{\tilde{\chi}_1^0} > 46
\mbox{ GeV} \right\}$.  The SM-like neutrinos mix with the neutralinos
and heavy neutrinos in a seesaw matrix.  We are able to reproduce the
correct neutrino mass scale 
but neglected the question of precise neutrino mass pattern\footnote{The issue of neutrino masses in 
the $\mu \nu$SSM was studied more extensively in 
Ref.~\cite{Ghosh:2008yh, Ghosh:2010zi}.}
 since any
desired neutrino mass pattern will not be difficult to achieve by
adjusting the small Yukawa couplings.  
Since we have already noted
that the smallness of the leptonic Yukawa couplings make their role in
the current SFOPT analysis insignificant, this does not present a
significant loss of generality.  

Because the squark, charged slepton, and left-handed sneutrino masses
are supported by their $\TeV$-scale SUSY-breaking mass parameters,
they are insensitive to parameters in the Higgs sector and are not
affected by the phenomenological lower bounds.  We compute the
mass spectrum of Higges and singlet sneutrinos at one-loop order using the
effective potential.  Since we choose the VEVs for these fields to be
real and there is no explicit CP-violation, the CP-even ($\phi_i$) and
CP-odd ($a_i$) components do not mix.  The mass matrices are given by
the curvature of the one-loop effective potential evaluated at the
zero temperature vacuum \footnote{Since the effective potential is
  defined as a sum over 1PI diagrams with zero external momentum, this
  definition of mass differs from the pole in the propagator by the
  difference of the scalar self-energy evaluated at $p =
  m_{\mathrm{pole}}$ and $p = 0$.  }
\begin{align}
	\left( M_{\phi}^2 \right)_{ij} = \left.  \frac{\partial^2 V_{\eff}^0}{\partial (\mathcal{R} \varphi_i) \partial (\mathcal{R} \varphi_j )} \right|_{\mathrm{VEV}}  && 
	\left( M_{a}^2 \right)_{ij} = \left.  \frac{\partial^2 V_{\eff}^0}{\partial (\mathcal{I} \varphi_i) \partial (\mathcal{I} \varphi_j )} \right|_{\mathrm{VEV}}  
\end{align}
where $\varphi_i \in \left\{ H_1^0, H_2^0, \snuc_j \right\}$.  We can separately impose the mass bounds $m_{\phi_1} > 92.8 \mbox{ GeV}$ and $m_{a_1} > 93.4 \mbox{ GeV}$.  
 
\begin{figure}[]
\begin{center}
\includegraphics[width=0.40 \textwidth]{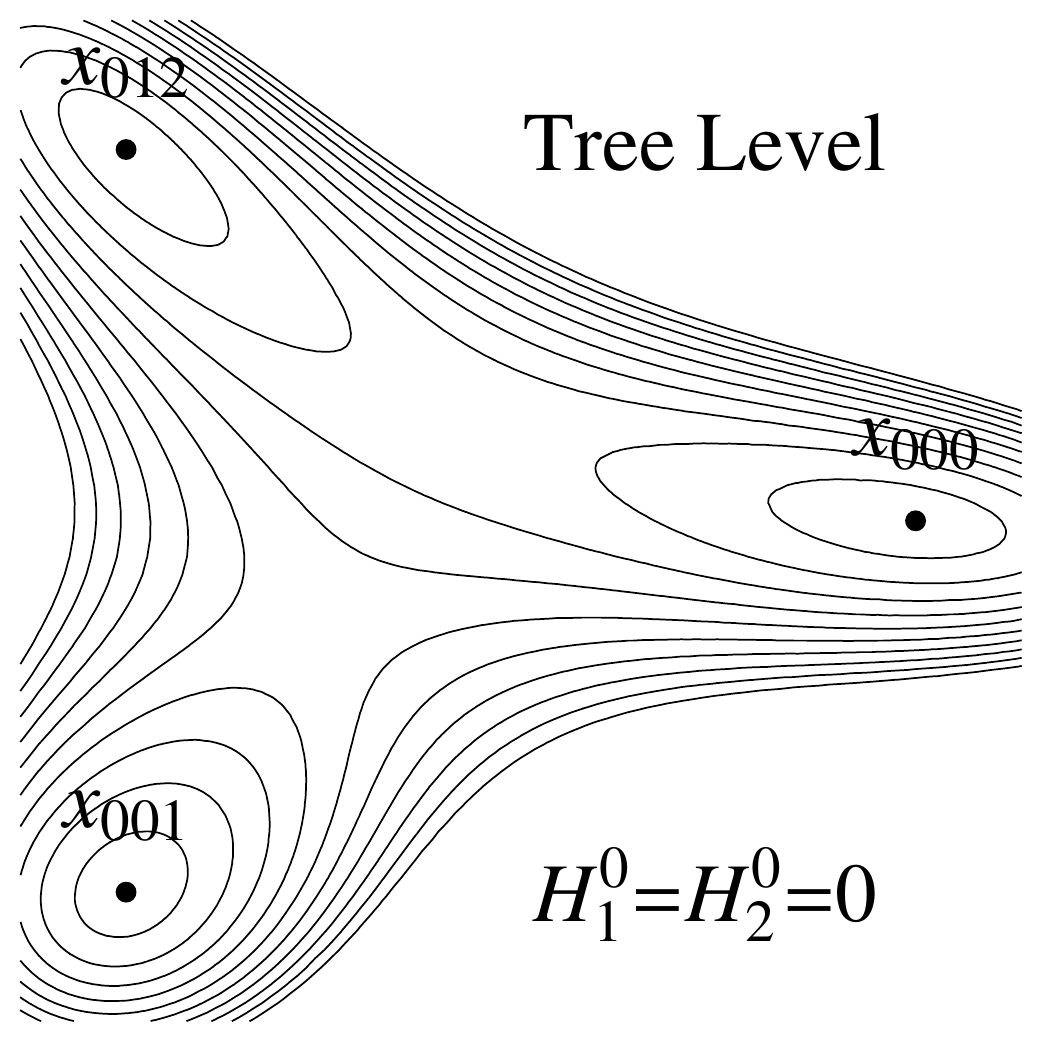} \hfill
\includegraphics[width=0.40 \textwidth]{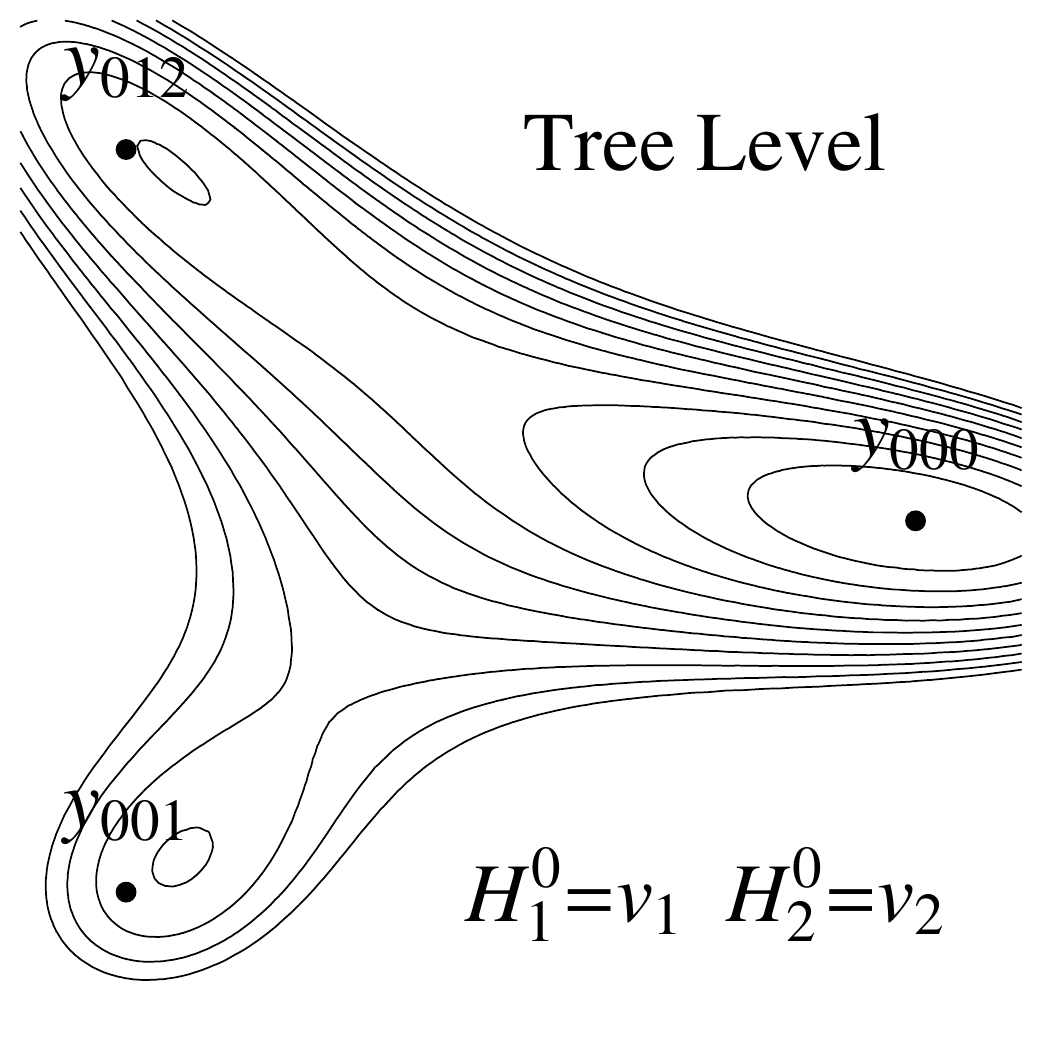}
\caption{\label{fig:FieldSpaceTree}
The tree level potential plotted over a slice of the
  $\snuc_i$ field space with $H_1^0 = H_2^0 = 0$ on the left and
  $H_1^0 / v_1 = H_2^0 /v_2 = 1$ on the right.  The labeled points are
  defined in the text, and a stationary point of the potential can be
  found at or near each of the labeled points.  The potential grows
  farther from the central region.  In the EW-preserving subspace, the
  three minima are degenerate, but the Higgs VEV selects out
  $\y_{000}$ as the global minimum.  }
\end{center}
\end{figure}

At each point in parameter space that satisfies the phenomenological mass bounds, we require the electroweak-breaking vacuum with $v(0) = 174 \GeV$ to be the global minimum of the one-loop effective potential.  This condition imposes particularly strong constraints on the parameter space.  To understand these constraints and the nature of our multistep phase transitions, we must discuss the structure of the $\left\{ H_1^0, H_2^0, \snuc_i \right\}$ field space and, in particular, determine the locations of the vacua that could potentially have lower energy than the EWSB vacuum.  Recall that in the subspace with $H_1^0 = H_2^0=0$ there is a $\mathbb{Z}_3 \otimes \mathbb{Z}_3 \otimes \mathbb{Z}_3$ symmetry at tree level given by \eref{eq:Z3independent}.  To locate the extrema in this field space we solve the three cubic equations
\begin{align}\label{eq:SingletMinEqn}
	\left. \frac{\partial V_0}{\partial \snuc_i} \right|_{H_1^0 = H_2^0 = 0}  = 0 \ \ \ \ \  i \in \left\{ 1, 2, 3 \right\}
\end{align}
for $\snuc_i$.  The solutions of \eref{eq:SingletMinEqn} which turn out to be minima (for our choice of the sign of $a_{\kappa}$) are given by 
\begin{align}
	\snuc_i &= \rho_i e^{i n_i \frac{2\pi}{3} } \ \ \ \ \ \ n_{i} \in \left\{ 0, 1, 2 \right\} \nonumber  \\
	\rho_i &= 0 \ \ \ \ \mathrm{or} \ \ \ \ \frac{1}{4 \kappa^2} \left( -a_{\kappa} + \sqrt{a_{\kappa}^2 - 8 m_{\snuc}^2 \kappa^2} \right) \approx v_{\snuc}
\end{align}
We will focus on the solutions with $\rho_1 = \rho_2 = \rho_3 \approx
v_{\snuc}$ because these minima are in general deeper than those with
$\rho_i = 0$.  Then, there are $3^3 = 27$ local minima in the $H_1^0 =
H_2^0 = 0$ field space, that we will refer to by $\x_{n_1 n_2 n_3}$
where the subscript indicates the phases of the three singlets.  The
$\left( \mathbb{Z}_3 \right)^3 \otimes \mathbb{S}_3 \otimes
\mathrm{CP}$ symmetry ensures the degeneracy of the 27
minima.  Moving away from $H_1^0 = H_2^0 = 0$ as illustrated in Fig.
\ref{fig:FieldSpaceTree}, the $\left( \mathbb{Z}_3 \right)^3$ symmetry
is broken to $\mathbb{Z}_3$ by terms in $V_0$ proportional to
$a_{\lambda}$ and $\lambda$.  We will use $\y_{n_1 n_2 n_3}$ to denote
the point in field space near to $\x_{n_1 n_2 n_3}$ but where $H_1^0 /
v_1 = H_2^0 / v_2 = 1$.  For example, in this notation, $\y_{000}$
corresponds to the EWSB vacuum in which the three singlets have real
VEVs.

\begin{figure}[]
\begin{center}
\includegraphics[width=0.40 \textwidth]{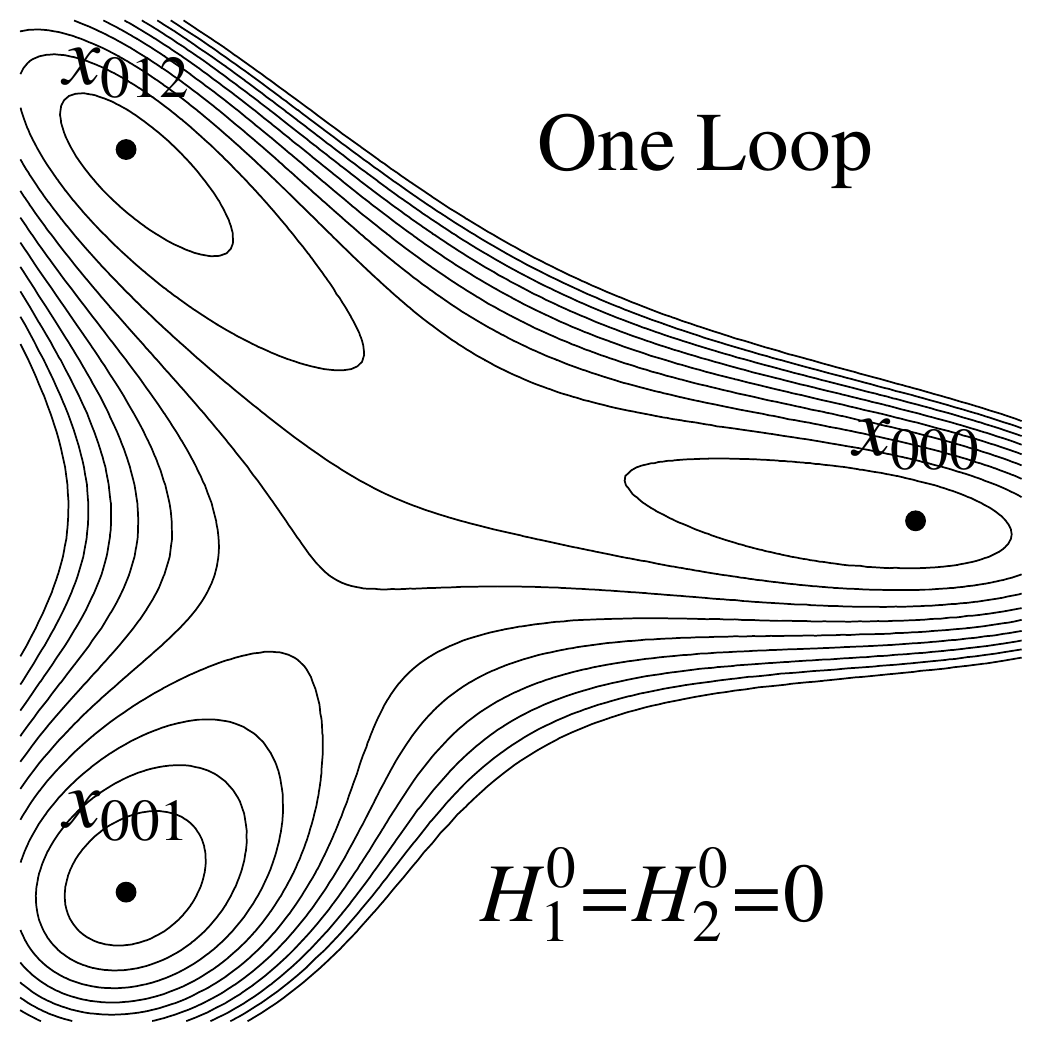} \hfill
\includegraphics[width=0.40 \textwidth]{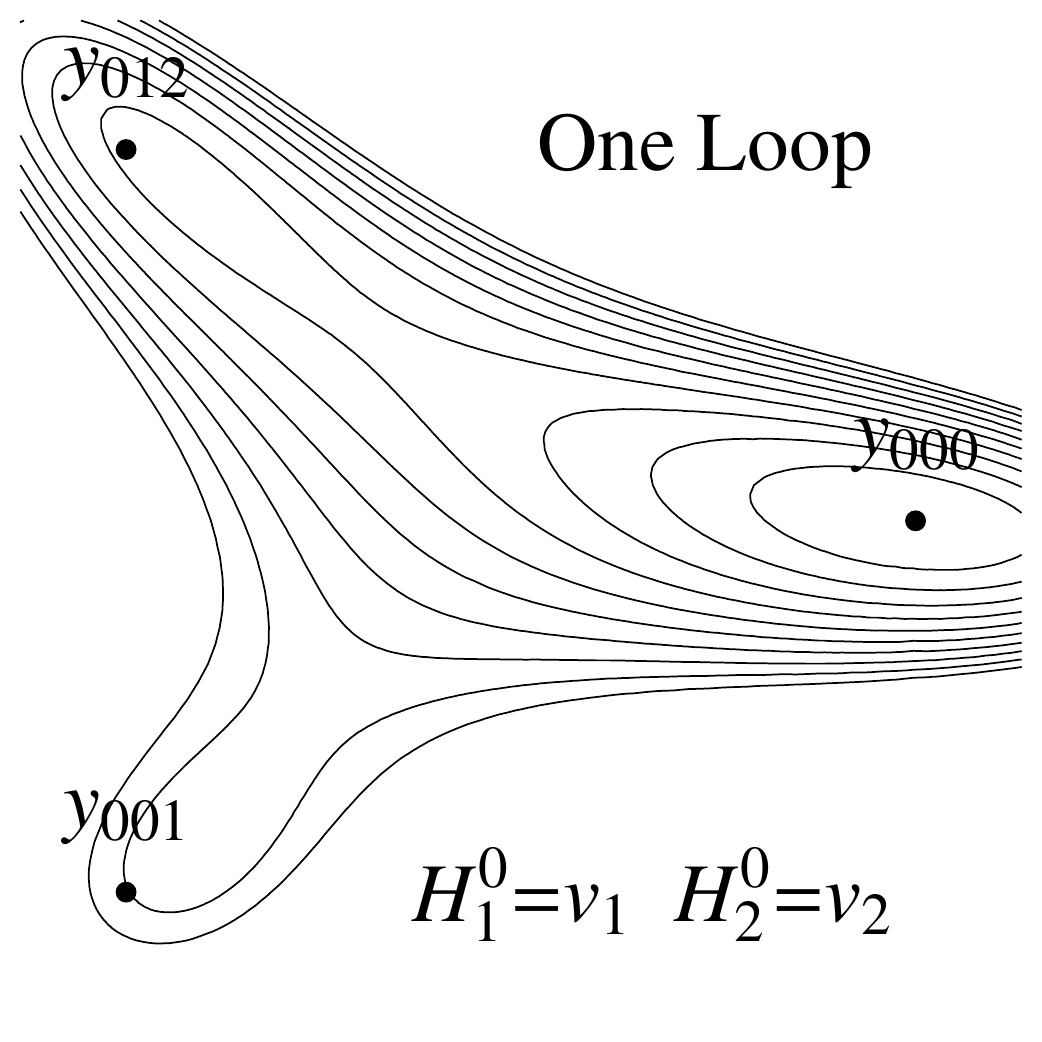}
\caption{\label{fig:FieldSpaceLoop}
Same as Fig.  \ref{fig:FieldSpaceTree} but the contours represent values of the one-loop effective potential at zero temperature.  The degeneracy is broken even in the EW-preserving subspace and induces $V_1^0 \left( \x_{012} \right) < V_1^0 \left( \x_{001} \right) < V_1^0 \left( \x_{000} \right)$ .  }
\end{center}
\end{figure}

At one-loop order, radiative corrections break the approximate $\left(
\mathbb{Z}_3 \right)^3$ symmetry [described above
Eq.~(\ref{eq:Z3independent})] and split the degeneracy of the $\x_{n_1
  n_2 n_3}$ minima as represented by Fig.  \ref{fig:FieldSpaceLoop}.
After including radiative corrections, the preserved symmetry group is
$ \mathbb{Z}_3 \otimes \mathbb{S}_3 \otimes \mathrm{CP}$.  (Here, as
an approximation, we are ignoring the fact that the subgroup
$\mathbb{S}_3 \otimes \mathrm{CP}$ is explicitly weakly broken in the
Lagrangian already while $\mathbb{Z}_3$ must be broken by
nonrenormalizable operators to evade cosmological inconsistencies
caused by domain walls.) The 27-fold degeneracy is split into three
classes: a 3-fold degeneracy of the points $\x_{i i i}$, a 6-fold
degeneracy of the points $\x_{i j k}$ for $i \neq j \neq k$, and a
18-fold degeneracy of the points $\x_{i i j}$ plus permutations for $i
\neq j$.  In order to discuss the phase transition we will choose one
representative from each class: $\x_{000}$, $\x_{012}$, and
$\x_{001}$.  In this notation, if we say a transition occurs from the
origin to $\x_{012}$ we mean that just below the critical temperature
the vacuum is localized nearby to one of the six field points in the
class that contains $\x_{012}$.

The radiative corrections will generally split the degeneracy in such
a way that some of the EW-preserving vacua will be depressed relative
to the EWSB vacuum and may cause the latter to become metastable.
This is both good and bad for the parameter space scan and phase
transition analysis.  It is bad because many points will be excluded
because the EWSB vacuum is only metastable.  On the other hand, it is
good because with appropriate tuning, we can obtain an EW-preserving
vacuum that is nearly degenerate with, but slightly higher than, the
EWSB vacuum.  Along the trajectory connecting these vacua, we can make
$E_{\eff} /\left(\lambda_{\eff} \phi(0)\right)$ arbitrarily close to
one half and obtain SFOPT.  In Appendix \ref{sec:derivingboundaries}
we include analytic bounds which must be satisfied to prevent the EWSB
vacuum from becoming metastable.

At each point in parameter space which satisfies the mass and vacuum
bounds described above, we calculate the critical temperature, $T_c$,
and Higgs VEV, $v ( T_c)$, at the electroweak phase transition.
The phase transition is calculated using the following procedure:
increase the temperature from zero in increments, at each temperature
minimize the thermal effective potential to find the EWSB vacuum
$\vec{v}_{EWSB}(T) = \left\{ v_1(T), v_2(T), v_{\snuc_i}(T) \right\}$,
also find whatever metastable vacua $\vec{v}_{\mathrm{MS},i}$ are near
to $\x_{n_1 n_2 n_3}$ and $\y_{n_1 n_2 n_3}$, as the temperature
increases the location and depth of these stationary points will
change, converge on the critical temperature $T_c$ at which the EWSB
vacuum becomes degenerate with one of the EW-preserving minima
$V_{\eff}^{T_c} \left( \vec{v}_{EWSB} (T_c) \right) = V_{\eff}^{T_c}
\left( \vec{v}_{\mathrm{MS},i_{\star}} (T_c) \right)$, compute $v
\left( T_c \right)= \sqrt{ \left| v_1 (T_c) \right|^2 + \left| v_2 (T_c) \right|^2}$ as the Higgs VEV in the broken phase.  Using this procedure, we obtain $T_c$ and $v
\left( T_c \right)$ for the lowest temperature phase transition.
Generally in this region of parameter space, multiple phase transition
steps are required to bring the field configuration from the
high-temperature symmetric phase to the zero temperature broken phase.
We must investigate separately earlier steps.

\begin{figure}[]
\begin{center}
\includegraphics[width=0.6 \textwidth]{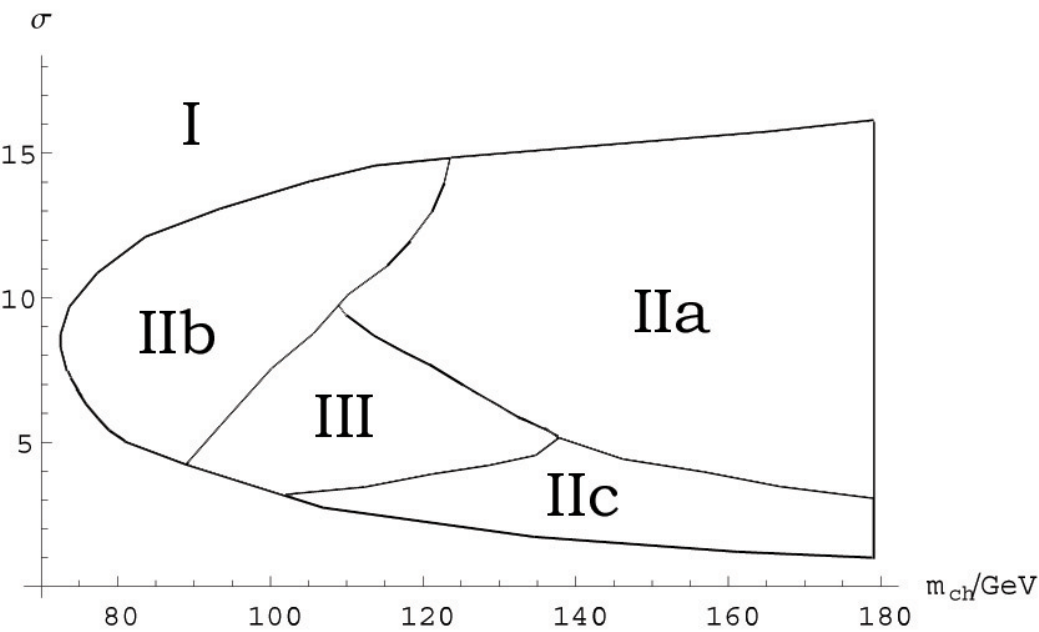} \hfill
\includegraphics[width=0.6 \textwidth]{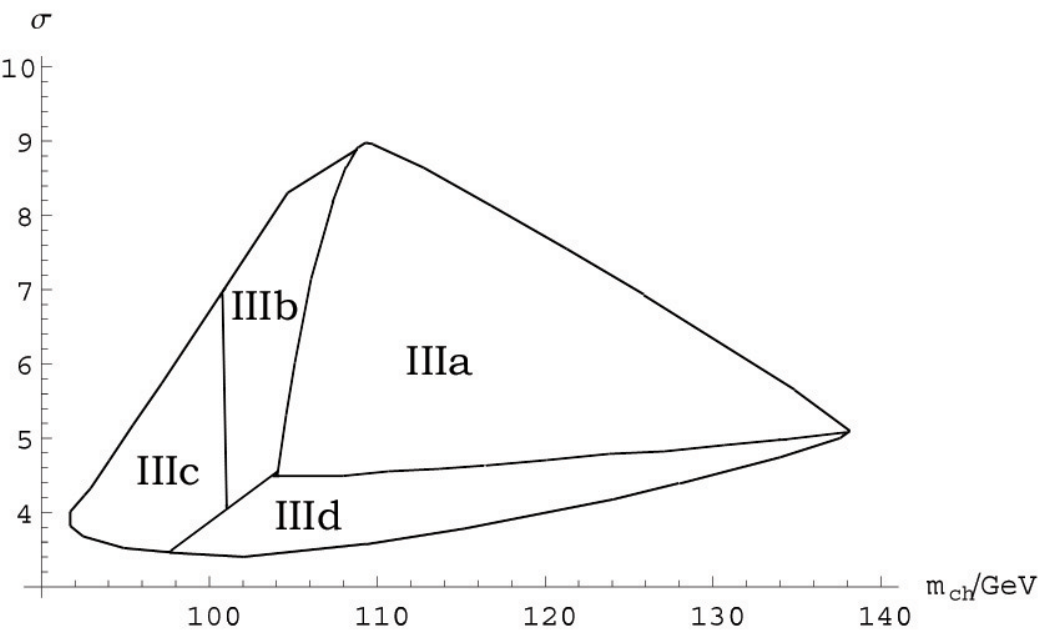}
\caption{
\label{fig:Scan1} A slice of the $\mu \nu$SSM parameter space.  Region I suffers from tachyons in the EWSB vacuum.  In region II the EWSB vacuum is metastable and we exclude these points.  In region III we calculate the electroweak phase transition and find that the path through field space can be classified into one of four types, shown on the right.  }
\end{center}
\end{figure}

The results of the 2000 point parameter space scan are summarized in
Fig.  \ref{fig:Scan1} where regions IIIa and IIIb are the only likely
viable regions for SFOPT EWBG.  We will describe the different regions
here and give an analytic derivation of the boundaries and their
parametric dependence in Appendix \ref{sec:derivingboundaries}.  The
points in region I are excluded because the EWSB vacuum, where $v (0)
= 174 \GeV$, contains a tachyonic direction.  The points in region II
are excluded because the EWSB vacuum is metastable.  For regions
IIa, IIb, and IIc, the actual vacuum can be found at the following points: the
origin of field space in region IIa, nearby to $x_{012}$ in region IIb, and
nearby to $y_{012}$ in region IIc.  That is, in regions IIa and IIb, the
electroweak phase transition does not occur.  Region IIc does not work
for EWBG as well as we will see below.  In region III there are no
tachyons, no false minima, and all phenomenological mass bounds are
satisfied, but as we will see only IIIa and IIIb are likely to give
acceptable phase transitions for EWBG.

The phase transition at each point can be classified into one of four
types based on the path that the vacuum follows through the $\left\{
H_1^0, H_2^0, \snuc_i \right\}$ field space.  In the largest region
IIIa, the PT makes two steps: from the origin to a $\left\{
\mathrm{EW}, \slashed{\mathrm{Z}_3}, \slashed{\mathbb{S}_3},
\slashed{\mathrm{CP}} \right\}$ phase and then to the $\left\{
\slashed{\mathrm{EW}}, \slashed{\mathrm{Z}_3}, \mathbb{S}_3,
\mathrm{CP} \right\}$ phase.  In region IIIb the EWPT occurs in one
step directly from the origin to the EW-broken phase.  In region IIIc,
the EW symmetry is broken by a second order phase transition in which
only $H_2^0$ gets a VEV; then, a first order phase transition occurs
giving the singlets VEVs.  Finally in region IIId the phase
transitions occur in three or four steps and there are multiple EWSB
phases, whose details for a representative point are discussed below.
However, as we will see, region IIId is unlikely to give an acceptable
of EWBG scenario.

To understand how the $\mu \nu$SSM phase transition differs from the NMSSM scenario, we have taken one representative parameter point from each sector of region III and followed the full phase transition from the origin $\x_O$ to the zero temperature EWSB vacuum $\y_{000}$.  In the tables, the minima above and below an arrow are degenerate at the temperature indicated.  A $0^+$ indicates that a second order phase transition occurs along the specified field direction.  

\begin{figure}[]
\begin{center}
\includegraphics[width=0.5 \textwidth]{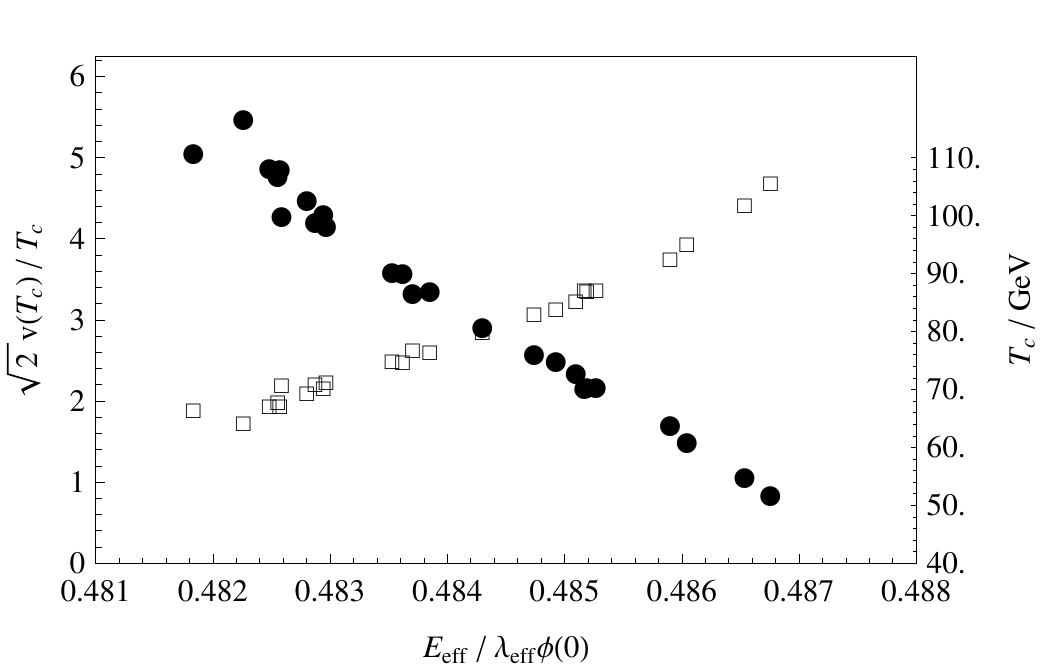}
\caption{\label{fig:Eoverlamv}
The order parameter (squares) and critical temperature (circles) plotted against $E_{\eff} / \lambda_{\eff} \phi \left( 0 \right)$ for the points in parametric region IIIb.  We calculate $E_{\eff} / \lambda_{\eff} \phi \left( 0 \right)$ using the tree level potential along the trajectory that connects the origin $\x_{O}$ and zero temperature vacuum $\y_{000}$.   }
\end{center}
\end{figure}

\begin{description}
	\item{\bf{IIIa.}} Two Step Transition via EW-preserving Phase:  $\x_O \xrightarrow{1PT} \x_{012} \xrightarrow{1PT} \y_{000}$\\
Representative point:  $\left\{ m_{\ch}, \sigma \right\} = \left\{108.8 \GeV, 9.12 \right\}$.  \\

\begin{table}[!h]
\begin{tabular}{|c|c|c|c|}
\hline
T & 
$\left\{ H_1^0, H_2^0, \snuc_i \right\}$ (GeV) & 
$ \sqrt{2} \frac{v(T)}{T}$ & 
Symmetries \\
\hline
\hline
$\gg M_w$ & 
$\left\{ 0, 0, 0, 0, 0 \right\}$	& 
0 & 
$\left\{ \mathrm{EW}, \mathbb{Z}_3, \mathbb{S}_3, \mathrm{CP} \right\}$ \\
$\xrightarrow{75.1 \GeV}$  & 
$ \left\{ 0, 0, 191.7, 191.7 e^{i 2\pi/3}, 191.7 e^{i 4 \pi/3} \right\}$ & 
0 & 
$\left\{ \mathrm{EW}, \slashed{\mathbb{Z}_3}, \slashed{\mathbb{S}_3}, \slashed{\mathrm{CP}} \right\}$ \\
\hline
 & 
$\left\{ 0, 0, 192.6, 192.6 e^{i 2\pi/3}, 192.6 e^{i 4 \pi/3} \right\}$	& 
0 & 
$\left\{ \mathrm{EW}, \slashed{\mathbb{Z}_3}, \slashed{\mathbb{S}_3}, \slashed{\mathrm{CP}} \right\}$ \\
$\xrightarrow{54.6 \GeV}$  & 
$ \left\{ 61.7, 160.7, 201.0, 201.0, 201.0 \right\}$ & 
4.46 & 
$\left\{ \slashed{\mathrm{EW}}, \slashed{\mathbb{Z}_3}, \mathbb{S}_3, \mathrm{CP} \right\}$ \\
\hline
$0 \GeV$  & 
$ \left\{ 62.5, 162.5, 201.5, 201.5, 201.5 \right\}$ & 
N/A & 
$\left\{ \slashed{\mathrm{EW}}, \slashed{\mathbb{Z}_3}, \mathbb{S}_3, \mathrm{CP} \right\}$\\
\hline
\end{tabular}
\caption{Phase transition path for the representative point in region IIIa. }
\label{tab:IIIaPath}
\end{table}%

At $T = 75.1 \GeV$, a first order phase transition gives the singlets
VEVs and breaks $\mathbb{Z}_3, \mathbb{S}_3$, and CP.  The EW symmetry
is broken by a strongly first order phase transition at $54.6$ GeV
which also restores $\mathbb{S}_3$ and CP.  Baryon number may be
generated at the strongly first order EW-breaking PT because
sphalerons are suppressed by $\sqrt{2} v (T_c) / T_c = 4.46$ inside
the bubble.

	\item{\bf{IIIb.}} One Step:  $\x_O \xrightarrow{1PT} \y_{000}$ \\
Representative point:  $\left\{ m_{\ch}, \sigma \right\} = \left\{102.5 \GeV, 7.22\right\}$.  \\

\begin{table}[!h]
\begin{tabular}{|c|c|c|c|}
\hline
T & 
$\left\{ H_1^0, H_2^0, \snuc_i \right\}$ & 
$ \sqrt{2} \frac{v(T)}{T}$ & 
Symmetries \\
\hline
\hline
$\gg M_w$ & 
$\left\{ 0, 0, 0, 0, 0 \right\}$	& 
0 & 
$\left\{ \mathrm{EW}, \mathbb{Z}_3, \mathbb{S}_3, \mathrm{CP} \right\}$ \\
$\xrightarrow{69.6 \GeV}$  & 
$ \left\{ 60.3, 157.7, 188.3, 188.3, 188.3 \right\}$ & 
3.43 & 
$\left\{\slashed{\mathrm{EW}}, \slashed{\mathbb{Z}_3}, \mathbb{S}_3, \mathrm{CP} \right\}$ \\
\hline
$0 \GeV$  & 
$ \left\{ 62.5, 162.5, 189.8, 189.8, 189.8 \right\}$ & 
N/A & 
$\left\{ \slashed{\mathrm{EW}}, \slashed{\mathbb{Z}_3}, \mathbb{S}_3, \mathrm{CP} \right\}$ \\
\hline
\end{tabular}
\caption{Phase transition path for the representative point in region IIIb.  }
\label{tab:IIIbPath}
\end{table}%

At $T = 69.6$ GeV, the Higgs and singlets obtain VEVs simultaneously
breaking the EW symmetry and $\mathbb{Z}_3$.  This one step phase
transition resembles the ones seen in certain parametric regions of
the NMSSM and other Higgs-singlet extensions.  A baryon number may be
generated since $\sqrt{2} v(T_c) / T_c = 3.43$ in the broken phase
will suppress washout.  For the parameters in region IIIb, we have
plotted in Fig.  \ref{fig:Eoverlamv} the order parameter and critical
temperature as functions of $E_{\eff} / \left( \lambda_{\eff} \phi(0)
\right)$ which we calculate using the tree level potential along the
trajectory that joins $\x_O$ and $\y_{000}$.  The order parameter
grows and the critical temperature decreases as $E_{\eff} /
\left(\lambda_{\eff} \phi ( 0)\right)$ approaches $1/2$ from below.
The data points do not extend all the way to $1/2$ because the
radiative corrections lift the potential in such a way that parameter
sets with $E_{\eff} /\left(\lambda_{\eff} \phi ( 0 )\right) \approx
1/2$ at tree level have a metastable EWSB vacuum at one-loop.

	\item{\bf{IIIc.}} Two Step via EWSB Phase:  $\x_O \xrightarrow{2PT} \y_{H_2} \xrightarrow{1PT} \y_{000}$\\
Representative point:  $\left\{ m_{\ch}, \sigma \right\} = \left\{95.1 \GeV, 5.37 \right\}$.  \\

\begin{table}[!h]
\begin{tabular}{|c|c|c|c|}
\hline
T & 
$\left\{ H_1^0, H_2^0, \snuc_i \right\}$ & 
$ \sqrt{2} \frac{v(T)}{T}$ & 
Symmetries \\
\hline
\hline
$\gg M_w$ & 
$\left\{ 0, 0, 0, 0, 0 \right\}$	& 
0 & 
$\left\{ \mathrm{EW}, \mathbb{Z}_3, \mathbb{S}_3, \mathrm{CP} \right\}$ \\
$\xrightarrow{116.4 \GeV}$  & 
$ \left\{ 0, 0^{+}, 0, 0, 0 \right\}$ & 
$0^+$ & 
$\left\{ \slashed{\mathrm{EW}}, \mathbb{Z}_3, \mathbb{S}_3, \mathrm{CP} \right\}$ \\
\hline
& 
$\left\{ 0, 101, 0, 0, 0 \right\}$	& 
2.5 & 
$\left\{ \slashed{\mathrm{EW}}, \slashed{\mathbb{Z}_3}, \mathbb{S}_3, \mathrm{CP} \right\}$ \\
$\xrightarrow{57.6 \GeV}$  & 
$ \left\{ 61.6, 160.6, 175.4, 175.4, 175.4 \right\}$ & 
4.25 & 
$\left\{ \slashed{\mathrm{EW}}, \slashed{\mathbb{Z}_3}, \mathbb{S}_3, \mathrm{CP} \right\}$ \\
\hline
$0 \GeV$  & 
$ \left\{ 62.5, 162.5, 176.2, 176.2, 176.2 \right\}$ & 
N/A & 
$\left\{ \slashed{\mathrm{EW}}, \slashed{\mathbb{Z}_3}, \mathbb{S}_3, \mathrm{CP} \right\}$ \\
\hline
\end{tabular}
\caption{Phase transition path for the representative point in region IIIc. }
\label{tab:IIIcPath}
\end{table}%

At a high temperature $T = 116.4$ GeV the EW symmetry is broken by a
second order phase transition along the up-type Higgs direction.  As
the temperature decreases, the global minimum of the effective
potential moves along the $H_2^0$ axis until it becomes degenerate
with a minimum localized near to $\y_{000}$.  A first order phase
transition occurs with $\sqrt{2} v(T_c) / T_c = 4.25$ inside the
bubble and $\sqrt{2} v(T_c) / T_c = 2.5$ outside the bubble.  In this
scenario, there is no baryon number generation.  Because the first
transition is of the second order, there is no coexistence of
phases.  The second transition is first order, but the sphaleron
transition rate is suppressed both inside and outside the bubble such
that $B+L$ is preserved on both sides.

	\item{\bf{IIId.}} Multi-Step:  $\x_O \xrightarrow{1PT}  \x_{012} \xrightarrow{1PT} \left( \y_{001} \mbox{ or } \y_{002} \right) \xrightarrow{1PT} \y_{000}$\\
Representative point:  $\left\{ m_{\ch}, \sigma \right\} = \left\{121.6 \GeV, 4.40 \right\}$.  \\

\begin{table}[!h]
\begin{tabular}{|c|c|c|c|}
\hline
T & 
$\left\{ H_1^0, H_2^0, \snuc_i \right\}$ & 
$ \sqrt{2} \frac{v(T)}{T}$ & 
Symmetries \\
\hline
\hline
$\gg M_w$ & 
$\left\{ 0, 0, 0, 0, 0 \right\}$	& 
0 & 
$\left\{ \mathrm{EW}, \mathbb{Z}_3, \mathbb{S}_3, \mathrm{CP} \right\}$ \\
$\xrightarrow{222.3 \GeV}$  & 
$ \left\{ 0, 0, 197, 197 e^{i 2 \pi / 3}, 197 e^{i 4 \pi/3} \right\} $ & 
0 & 
$\left\{ \mathrm{EW}, \slashed{\mathbb{Z}_3}, \slashed{\mathbb{S}_3}, \slashed{\mathrm{CP}} \right\}$ \\
\hline
 & 
$\left\{ 0, 0, 229, 229 e^{i 2 \pi/3}, 229 e^{i 4 \pi/3} \right\}$	& 
0 & 
$\left\{ \mathrm{EW}, \slashed{\mathbb{Z}_3}, \slashed{\mathbb{S}_3}, \slashed{\mathrm{CP}} \right\}$ \\
$\xrightarrow{128 \GeV}$  & 
$ \left\{ 0^{+}, 0^{+}, 229, 229 e^{i 2\pi/3}, 229 e^{i 4 \pi/3} \right\}$ & 
$0^+$ & 
$\left\{ \slashed{\mathrm{EW}}, \slashed{\mathbb{Z}_3}, \slashed{\mathbb{S}_3}, \slashed{\mathrm{CP}} \right\}$ \\
\hline
 & 
$ \left\{ 0^+, 82, 232, 232 e^{i 2 \pi/3}, 232 e^{i 4 \pi/3} \right\}$ & 
1.1 & 
$\left\{\slashed{\mathrm{EW}}, \slashed{\mathbb{Z}_3}, \slashed{\mathbb{S}_3}, \slashed{\mathrm{CP}}\right\}$ \\
$\xrightarrow{105 \GeV}$  & 
$ \left\{ 29 e^{i 1.9 \pi}, 117 e^{i 1.9 \pi}, 229, 229, 232 e^{i 2 \pi/3}  \right\}$ & 
1.6 & 
$\left\{\slashed{\mathrm{EW}}, \slashed{\mathbb{Z}_3}, \slashed{\mathbb{S}_3}, \slashed{\mathrm{CP}} \right\}$ \\
\hline
 & 
$ \left\{ 32.0, 128.0, 230.3, 230.3, 232.7 e^{i 2 \pi/3} \right\}$ & 
2.10 & 
$\left\{\slashed{\mathrm{EW}}, \slashed{\mathbb{Z}_3}, \slashed{\mathbb{S}_3}, \slashed{\mathrm{CP}} \right\}$ \\
$\xrightarrow{89.0 \GeV}$  & 
$ \left\{ 58.4, 152.3, 224.4, 224.4, 224.4  \right\}$ & 
2.59 & 
$\left\{ \slashed{\mathrm{EW}}, \slashed{\mathbb{Z}_3}, \mathbb{S}_3, \mathrm{CP} \right\}$ \\
\hline
$0 \GeV$  & 
$ \left\{ 62.5, 162.5, 225.1, 225.1, 225.1 \right\}$ & 
N/A & 
$\left\{ \slashed{\mathrm{EW}}, \slashed{\mathbb{Z}_3}, \mathbb{S}_3, \mathrm{CP} \right\}$ \\
\hline
\end{tabular}
\caption{Phase transition path for the representative point in region IIId. }
\label{tab:IIIdPath}
\end{table}%

At this parametric point, the phase transition occurs in four steps
with the EW symmetry broken in the second step by a second order phase
transition.  As the temperature drops from 128 GeV to 105 GeV, the
sphaleron becomes increasingly suppressed.  When the 1PT occurs at 105
GeV, the sphaleron is inactive, such that there will be no B-number
generation.  Not every phase transition in region IIId follows this
particular PT path, but the PTs are generally multistep with at least
one EWSB intermediate phase and transitional CP violation.

\end{description}

To give an impression of the particle masses in this region of parameter space, we include here the spectrum for the representative point in region IIIb where  $\left\{ m_{\ch}, \sigma \right\} = \left\{102.5 \GeV, 7.22\right\}$.  The slepton, squark, gaugino, and left-handed sneutrino masses are all $\ord{\mathrm{TeV}}$ because we have fixed the soft masses in these sectors at a fiducial SUSY-breaking scale.  In the case of the $\snu$, we solve for the soft mass using the minimization equations to find  $m_{\tilde{L}}^2 \approx -a_{\nu} v_{\snuc} v_2 / v_{\snu} $, yet it is still typically TeV scale because $A_{\nu} = a_{\nu} / Y_{\nu} = -1 \TeV$.  The remaining fermion and charged scalar spectrum is given by 
\begin{align}
	m_{H^{\pm}} & = 342 \GeV \nonumber \\
	m_{\tilde{H}^{\pm}} & = 96 \GeV \nonumber \\
	m_{\nu^c} & = 260, 243, 243 \GeV \nonumber \\
	m_{\tilde{\chi}_{1,2}^0} & = 88.7, 94.5 \GeV, \mbox{   mostly Higgsino} \nonumber \\
	m_{\nu} & = 41, 41, 7.8 \mathrm{\ meV} 
\end{align}
which are calculated at tree level.  The LSP is a Higgsino with mass $88.7 \GeV$.  The degeneracies present in the neutrino sector result from the $\mathbb{S}_3$ symmetry of our Lagrangian.  By allowing the left-handed sneutrinos to have different VEVs or choosing different values for the $Y_{\nu}$ Yukawas, we could obtain a correct neutrino hierarchy.  We include these masses here to demonstrate that the seesaw matrix produces the correct mass scale for the light neutrinos.  The neutral scalar masses are calculated at one-loop using the effective potential.  Because there is significant mixing, we have included their mass eigenvalues and field composition in Table \ref{tab:NeutralScalarSpec}.  Once again the degeneracies are a result of our $\mathbb{S}_3$ symmetry in the singlet sector.  The lightest Higgs is mostly up-type with a mass of 110 GeV at this parametric point and only varies by $10 \GeV$ over all of region III.  

\begin{table}[h]
\begin{tabular}{|c|c|c|c|c|c|}
\hline
 Mass (GeV) & $\Re H_1^0 $ & $\Re H_2^0$ & $\Re \snuc_1$ & $\Re \snuc_2$ & $\Re \snuc_3$ \\
\hline
395		& 0.83 	& 0.16	& 0.00	& 0.00	& 0.00 \\
140		& 0.01	& 0.21	& 0.26	& 0.26	& 0.26 \\
128		& 0.00	& 0.00	& 0.58	& 0.01	& 0.41 \\
128		& 0.00	& 0.00	& 0.09	& 0.65	& 0.07 \\
110 		& 0.16	& 0.63	& 0.07	& 0.07	& 0.07 \\
\hline
\end{tabular}
\hfill
\begin{tabular}{|c|c|c|c|c|c|}
\hline
 Mass (GeV) & $\Im H_1^0 $ & $\Im H_2^0$ & $\Im \snuc_1$ & $\Im \snuc_2$ & $\Im \snuc_3$ \\
\hline
439		& 0.48 	& 0.07	& 0.15	& 0.15	& 0.15 \\
369		& 0.00	& 0.00	& 0.02	& 0.39	& 0.59 \\
369		& 0.00	& 0.00	& 0.65	& 0.27	& 0.08 \\
314		& 0.39	& 0.06	& 0.18	& 0.18	& 0.18 \\
\hline
\end{tabular}
\label{tab:NeutralScalarSpec}
\caption{CP-even and CP-odd Higgs masses and mixings for a sample parameter point.  The field composition is described by the squared eigenvector associated with each eigenvalue.  }
\end{table}

\section{\label{sec:domainwall}Domain Walls}
It is well known
\cite{Abel:1995wk,Vilenkin:1984ib,Ellis:1986mq,Gelmini:1988sf,Rai:1992xw,Abel:1995uc}
that domain wall formation can be cosmologically problematic when
spontaneous breaking of discrete symmetry occurs.  In our scenario, we
have only one ``exact'' discrete symmetry $\mathbb{Z}_3$ at the level
of explicit parametrization of the Lagrangian.  Because of the
undesirable cosmological consequences of domain walls, we have
implicitly assumed that this symmetry is broken by nonrenormalizable
operators which are cutoff by a scale \footnote{Because we have
  $\lambda^2+\kappa^2 <0.5$ in the parametric regime of interest, the
  couplings should remain perturbative up to close to the GUT scale
  \cite{Miller:2003ay} (see e.g. \cite{Escudero:2008jg} for explicit
  plots which suggest that our parametric choice is close to the
  border of perturbativity up to the GUT scale).  Thus we are not
  severely restricted in the cutoff scale of our effective field
  theory.} larger than many TeV (such as not to disrupt the effective
potential analysis).  In addition, we have approximate discrete
symmetries such as $\mathbb{S}_3\otimes \mathrm{CP}$ which a priori can cause
problems if the symmetry breaking operators are overly suppressed.
However, the set of electroweak symmetry breaking vacua of interest in
this paper does not break $\mathbb{S}_3\otimes \mathrm{CP}$ (i.e. our symmetry
breaking pattern can naturally select a $\mathbb{S}_3\otimes \mathrm{CP}$
singlet VEV to be the lowest energy vacuum as partly demonstrated in
Appendix \ref{sec:slection}).  Hence, we will neglect any transient
behavior and focus on $\mathbb{Z}_3$ domain walls even though the
analysis is not very specific to the discrete group.\footnote{For
  example, if one wanted to analyze transient domain walls associated
  with $\mathbb{S}_3$ breaking, one can easily work out from our
  Lagrangian the leading effective scalar operator breaking
  $\mathbb{S}_3$ and use the result at the end of this section.}
Although a full analysis of domain wall histories is beyond the scope
of this paper, here we briefly estimate the effects of the suppressed
symmetry breaking operators that will alleviate the cosmological
problems associated with domain walls that may form when the discrete
symmetries considered in this paper are spontaneously broken.  We will
follow closely the work of Ref.~\cite{Abel:1995wk}.

In Ref.~\cite{Abel:1995wk}, it is estimated that during the
approximate discrete symmetry breaking phase transition, domain
walls separating approximately degenerate minima are formed.  Then a
simplified model of domain wall evolution is considered which
approximately accounts for the surface tension of the bubble, the
friction coming from bubble wall interaction with the plasma, and the
pressure coming from energy density difference between the
approximately degenerate minima.  This last ingredient (pressure from
energy density difference) is what will be coming from the inclusion
of suppressed symmetry breaking operators, and we will refer to this
simply as ``pressure difference.''  If the pressure difference dominates,
one of the approximately degenerate phases will eat away at the higher
energy phase regions and eventually dominate in a time scale
controlled by the strength of the symmetry breaking operator.  

Estimating the friction to be negligible, an approximate sufficient
condition for curing the possible domain wall problem from a
cosmological perspective is to have the pressure difference dominate
before the equilibrium initial condition period of big bang
nucleosynthesis: i.e. before the photon temperature reaches about $10$
MeV.  Explicitly, assuming order unity Lorentz factor $\gamma$ for the
bubble wall speed, one must require
\begin{equation}
\epsilon > \frac{\sigma}{R(t)}
\end{equation}  
where $\epsilon$ is the energy density difference coming from
suppressed symmetry breaking operators, $\sigma$ is the energy per
unit area of the bubble wall, and $R(t)$ is the time dependent radius of
a typical bubble.  For a dimension $4+u$ nonderivative operator
consisting of scalars only, $\epsilon$ can be estimated as
\begin{equation}
\epsilon \sim c_u \frac{v^{4+u}}{\Lambda^u}
\end{equation}
where $\Lambda$ is the cutoff scale and we have assumed all scalar
VEVs to be of common order $v\equiv 174$ GeV (which is appropriate for
our scenario). To be able to treat $u=0$, we will set $\Lambda = 100$
TeV and find a bound on the value of $c_u$ for different values of
$u$.  Assuming $R \sim t \sim 1/H$ (where $H$ is the Hubble expansion
rate) and $\sigma \sim v^3$, we find
\begin{equation}
c_u > 10^{-24} 500^u
\end{equation}
for $u \ge 0$.\footnote{This result can easily be checked to be
  consistent with Ref.~\cite{Abel:1995wk} for $u=1$.}  Hence, as long
as the cutoff is not required to be very large (in contrast with the
assumption of Ref.~\cite{Abel:1995wk}) or the accidental symmetry arising
from the UV completion quantum numbers do not make $u$ too large, this
bound is very easy to satisfy for the $\mathbb{Z}_3$ domain wall
problem.  Of course, if the cutoff is taken to be high and/or a UV
completion is desired without fine tuning, model building challenges
along the lines of Refs.~\cite{Abel:1996cr,Panagiotakopoulos:1998yw}
exist.

\section{Summary}
We have uncovered a $\mu\nu$SSM parametric region giving rise to a
first order phase transition sufficiently strong to be useful for the
electroweak baryogenesis scenarios involving electroweak symmetry
breaking bubbles as the source of out of equilibrium and $SU(2)_L$
$F\tilde{F}$ operators as a source of baryon number violation.  The
parametric region corresponds to tuning the soft terms in the
Lagrangian $a_{\lambda}H_{1} \! \cdot \!  H_{2}\tilde{\nu}_{i}^{c}$
and $-m_{\tilde{\nu}^{c}}^{2}|\tilde{\nu}_{i}^{c}|^{2}$ to achieve
Eq.~(\ref{eq:crucialsaturate}).  The numerical values of the uncovered
parametric region is in the paragraph containing
Eq.~(\ref{eq:sigmadef}) and regions IIIa and IIIb depicted in
Fig.~\ref{fig:Scan1}.  As expected, the Yukawa coupling of the
singlets to the leptonic sector does not play a role in determining
the strength of the phase transitions because of the weakness of the
coupling tied to the smallness of the neutrino masses.

The region IIIa transitions are two-step transitions in which the
electroweak symmetry breaking is the second transition that starts
from a phase in which the singlet scalars of the $\mu\nu$SSM have a
nonzero vacuum expectation value (e.g. starts from a vacuum which
spontaneously breaks the approximate $S_3$ symmetry in the singlet
sector).  These transitions contain a rotation in the singlet field
space and do not have an analog in the NMSSM transitions because of
the different dimensionality in the singlet complex vector space.  The
region IIIb transitions are the ones in which electroweak symmetry
breaking transition starts from the origin of the scalar field space.
All these transitions have useful descriptions in terms of the
representations of the approximate discrete symmetries in the system.

Our phenomenological bounds were rather minimal and placed using
Ref.~\cite{PDG}, but in many parametric regions, the observables are
sufficiently far away from the bounds that the plausibility of the
phenomenological self-consistency is strong.  Follow-up possibilities
include a more complete collider related phenomenological
investigation in this parametric regime, studies of domain wall
histories due to weak global symmetry breaking operators, and a
complete computation of CP asymmetry creation and transport leading
to baryon asymmetry.

Given that the $\mu\nu$SSM had to give up the popular thermal leptogenesis
scenario due to its low scale implementation of the type I seesaw,
this work is of interest as it shows that electroweak baryogenesis may
be a promising avenue to create baryon asymmetry in this class of
models.  Given that the $\mu\nu$SSM is one of the few supersymmetric
models in which all dynamical degrees of freedom responsible for the
neutrino mass may be accessible at TeV-scale colliders, it is
encouraging that the model has a good chance at being consistent with
the observed baryon asymmetry in the universe.

\section*{Acknowledgments}
We thank Amjad Ashoorioon, Lisa Everett, Aki Hashimoto, Carlos
Mu\~{n}oz, David Morrissey, and Michael Ramsey-Musolf for useful
correspondence.  We thank the hospitality of KIAS where part of this
work was completed.  This work is supported by the DOE through grant
DE-FG02-95ER40896.

\appendix

\section{\label{sec:massmatrices}Field-Dependent Mass Matrices}

Here we include the tree level, field-dependent mass matrices which
are required to compute the one-loop radiative corrections.  We fix
the charged scalars at their vanishing VEVs, let the left-handed
sneutrino VEV $\snu_i = v_{\snu}$ be real, and treat the matrices as
functions of the complex fields $\left\{ H_1^0, H_2^0, \snuc_i
\right\}$.  \\


\textbf{Neutralinos.}  Using the basis $ \left( \chi^0 \right)^T =
\left\{ \tilde{B}, \tilde{W}_3, \tilde{H}_1^0, \tilde{H}_2^0, \nu_1^c,
\nu_2^c, \nu_3^c, \nu_1, \nu_2, \nu_3 \right\}$, the mass term appears
as $\mathcal{L} \ni -\frac{1}{2} \left( \chi^0 \right)^T M_{\chi^0}
\left( \chi^0 \right) + \mathrm{h.c.}$ with $n_{\chi^0} = -2$ and
\begin{align}
	M_{\chi^0} & = 
\begin{pmatrix}
\mathcal{M} & m \\
m^T & 0_{3 \times 3}
\end{pmatrix} \\
	m^T & = 
\begin{pmatrix}
-\frac{g_1}{\sqrt{2}} v_{\snu} & \frac{g_2}{\sqrt{2}} v_{\snu} & 0 & Y_{\nu} \snuc_1 & Y_{\nu} H_2^0 & 0 & 0 \\
-\frac{g_1}{\sqrt{2}} v_{\snu} & \frac{g_2}{\sqrt{2}} v_{\snu} & 0 & Y_{\nu} \snuc_2 & 0 & Y_{\nu} H_2^0  & 0 \\
-\frac{g_1}{\sqrt{2}} v_{\snu} & \frac{g_2}{\sqrt{2}} v_{\snu} & 0 & Y_{\nu} \snuc_3 & 0 & 0 & Y_{\nu} H_2^0  
\end{pmatrix}
\end{align}
and $\mathcal{M}$ is a symmetric, sparse matrix with nonzero elements
\begin{align}
\mathcal{M}_{\tilde{B} \tilde{B}} & = M_1 \\
\mathcal{M}_{\tilde{B} \tilde{H}_1^0} & = - \frac{g_1}{\sqrt{2}} \left( H_1^0 \right)^{\ast} \\
\mathcal{M}_{\tilde{B} \tilde{H}_2^0} & =  \frac{g_1}{\sqrt{2}} \left( H_2^0 \right)^{\ast} \\
\mathcal{M}_{\tilde{W}_3 \tilde{W}_3} & = M_2 \\
\mathcal{M}_{\tilde{W}_3 \tilde{H}_1^0} & = \frac{g_2}{\sqrt{2}} \left( H_1^0 \right)^{\ast} \\
\mathcal{M}_{\tilde{W}_3 \tilde{H}_2^0} & = - \frac{g_2}{\sqrt{2}} \left( H_2^0 \right)^{\ast} \\
\mathcal{M}_{\tilde{H}_1^0 \tilde{H}_2^0} & = - \lambda \left( \snuc_1 + \snuc_2 + \snuc_3 \right) \\
\mathcal{M}_{\tilde{H}_1^0 \nu_i^c } & = - \lambda H_2^0 \\
\mathcal{M}_{\tilde{H}_2^0 \nu_i^c } & = - \lambda H_1^0 + Y_{\nu} v_{\snu} \\
\mathcal{M}_{\nu_i^c \nu_j^c } & = 2 \kappa \snuc_i \delta_{ij}~~.
\end{align}

\textbf{Charginos.}  Using the basis $\left( \Psi^+ \right)^T = \left\{ -i \tilde{\lambda}^+, \tilde{H}_2^+, e_R^+, \mu_R^+, \tau_R^+ \right\}$ and $\left( \Psi^- \right)^T = \left\{ -i \tilde{\lambda}^-, \tilde{H}_1^-, e_L^-, \mu_L^-, \tau_L^- \right\}$ the mass term appears as $\mathcal{L} \ni - \frac{1}{2} \left( \Psi^- \right)^T M_{\chi^{\pm}} \left( \Psi^+ \right) + \mbox{ transpose}$ 
with $n_{\chi^{\pm}} = -2$ and
\begin{align}
	M_{\chi^{\pm}} = 
\begin{pmatrix}
M_2 & g_2 \left( H_2^0 \right)^{\ast} & 0 & 0 & 0 \\
g_2 \left( H_1^0 \right)^{\ast} & \lambda \left( \snuc_1 + \snuc_2 + \snuc_3 \right) & - Y_e v_{\snu} & - Y_{\mu} v_{\snu} & - Y_{\tau} v_{\snu} \\
g_2 v_{\snu} & - Y_{\nu} \snuc_1 & Y_e H_1^0 & 0 & 0 \\
g_2 v_{\snu} & - Y_{\nu} \snuc_2 & 0 & Y_{\mu} H_1^0  & 0 \\
g_2 v_{\snu} & - Y_{\nu} \snuc_3 & 0 & 0 & Y_{\tau} H_1^0
\end{pmatrix} .
\end{align}

\textbf{Gauge Bosons.}  The propagators and field-dependent masses in the gauge sector have gauge dependence.  We work in the Landau gauge ($\xi = 0$), in which the scalar component and ghost propagators have no field dependence.  The charged gauge bosons have field-dependent mass 
\begin{align}
	M_{W^{\pm}}^2 =  \frac{g_2^2}{2} \left( \left| H_1^0 \right|^2 + \left| H_2^0 \right|^2 + 3 v_{\snu}^2 \right) 
\end{align}
and in the basis $\left\{ W_3, B \right\}$ the neutral gauge bosons have the mass matrix 
\begin{align}
	M_{W_3 B}^2 = 
\begin{pmatrix}
\frac{g_2^2}{2} \left( \left| H_1^0 \right|^2 + \left| H_2^0 \right|^2 + 3 v_{\snu}^2 \right)  & -\frac{g_1 g_2}{2}  \left( \left| H_1^0 \right|^2 + \left| H_2^0 \right|^2 + 3 v_{\snu}^2 \right)   \\
-\frac{g_1 g_2}{2}  \left( \left| H_1^0 \right|^2 + \left| H_2^0 \right|^2 + 3 v_{\snu}^2 \right) & \frac{g_1^2}{2} \left( \left| H_1^0 \right|^2 + \left| H_2^0 \right|^2 + 3 v_{\snu}^2 \right)
\end{pmatrix} .
\end{align}
In order to count the degrees of freedom in the gauge sector, we must
distinguish longitudinal and transverse components of the gauge boson
fields, $2 n_{W^{\pm}_L} = n_{W^{\pm}_T} = 4$ and $2 n_{W_3 B_L} =
n_{W_3 B_T} = 2$.  We do this because only the longitudinal components
receive thermal mass corrections in the computation of the daisy
correction \eref{eq:Vdaisy}, \cite{Carrington:1991hz}.  Note that $v_{\snu}^2$ is numerically
negligible in these equations over all parameter regions of
interest.

\textbf{Squarks.}  Because we assume there is no intergenerational mixing in
the squark sector, the squark mass matrix block diagonalizes.  The
$i^{\mathrm{th}}$ generation up- and down-type squarks have mass terms
$\mathcal{L} \ni - \frac{1}{2} \tilde{q}_i^{\dagger} M_{\tilde{q}_i}^2
\tilde{q}_i$ in the basis $\tilde{q}_i = \left\{ \tilde{q}_{L_i} ,
\tilde{q}_{R_i}^{\ast} \right\}$ with $n_q = 12$ and
\begin{align}
\left( M_{\tilde{u}_i}^2 \right)_{11} & = M_{Q}^2 + \frac{1}{6} \left( \frac{3g_2^2}{2} - \frac{g_1^2}{2} \right) \left( \left| H_1^0 \right|^2 - \left| H_2^0 \right|^2 + 3 v_{\snu}^2 \right) + Y_{u_i}^2 \left|H_2^0 \right|^2 \\
\left( M_{\tilde{u}_i}^2 \right)_{12} & =\left( M_{\tilde{u}_i}^2 \right)^{\ast}_{21} = a_u \left( H_2^0 \right)^{\ast} - Y_{\nu} Y_{u} v_{\snu} \left( \snuc_1 + \snuc_2 + \snuc_3 \right) - Y_{u_i} \lambda \left( H_1^0 \right) \left( \snuc_1 + \snuc_2 + \snuc_3 \right) \\
\left( M_{\tilde{u}_i}^2 \right)_{22} & = m_{\tilde{u}^c}^2 + \frac{g_1^2}{3} \left( \left| H_1^0 \right|^2 - \left| H_2^0 \right|^2 + 3 v_{\snu}^2 \right) + Y_{u_i}^2 \left|H_2^0 \right|^2 \\
\left( M_{\tilde{d}_i}^2 \right)_{11} & = M_{Q}^2 - \frac{1}{6} \left( \frac{3g_2^2}{2} + \frac{g_1^2}{2} \right) \left( \left| H_1^0 \right|^2 - \left| H_2^0 \right|^2 + 3 v_{\snu}^2 \right) + Y_{d_i}^2 \left| H_1^0 \right|^2 \\
\left( M_{\tilde{d}_i}^2 \right)_{12} & = \left( M_{\tilde{d}_i}^2 \right)^{\ast}_{21} = a_d \left( H_1^0 \right) - Y_{d_i} \lambda \left( H_2^0 \right)^{\ast} \left( \snuc_1 + \snuc_2 + \snuc_3 \right)^{\ast} \\
\left( M_{\tilde{d}_i}^2 \right)_{22} & = m_{\tilde{d}^c}^2 - \frac{g_1^2}{6} \left( \left| H_1^0 \right|^2 - \left| H_2^0 \right|^2 + 3 v_{\snu}^2 \right) + Y_{d_i}^2 \left|H_1^0 \right|^2~~.
\end{align} 

\textbf{Charged Scalars.}  The charged Higgs mixes with the charged sleptons.  Using the basis $S^+ = \left\{ H_1^+, H_2^+, \tilde{e}_L^+, \tilde{e}_R^+, \tilde{\mu}_L^+, \tilde{\mu}_R^+, \tilde{\tau}_L^+, \tilde{\tau}_R^+ \right\}$, the mass term is $\mathcal{L} \ni - S^+ M_{H^{\pm}}^2 S^-$ with $n_{H^{\pm}} = 2$ and the elements of the Hermitian mass matrix are
\begin{align}
	\left( M_{H^{\pm}}^2 \right)_{H_1 H_1} &= 
	m_{H_1}^2 
	+ v_{\snu}^2 \sum_{i=1}^3 Y_{e_i}^2 
	+ \frac{1}{4} \left( g_1^2 + g_2^2 \right) \left| H_1^0 \right|^2 
	+ \frac{1}{4} \left( g_1^2 - g_2^2 \right) \left( 3 v_{\snu}^2 - \left| H_2^0 \right|^2  \right) \nonumber\\
& ~~~~	+ \lambda^2 \left| \snuc_1 + \snuc_2 + \snuc_3 \right|^2 \\
	\left( M_{H^{\pm}}^2 \right)_{H_1 H_2} &= 
	a_{\lambda} \sum_{i=1}^3 \snuc_i 
	+ 3 v_{\snu} Y_{\nu} \lambda \left( H_2^0 \right)^{\ast} 
	+ \left( \frac{1}{2} g_2^2 -3 \lambda^2 \right) \left( H_1^0 H_2^0 \right)^{\ast}  
	+ \kappa \lambda \sum_{i=1}^3 \left( \snuc_i \right)^{2 \ast} \\
	\left( M_{H^{\pm}}^2 \right)_{H_2 H_2} &= 
	m_{H_2}^2 
	+ \frac{1}{4} \left( g_1^2 + g_2^2 \right) \left| H_2^0 \right|^2 
	- \frac{1}{4} \left( g_1^2 - g_2^2 \right) \left( 3 v_{\snu}^2 + \left| H_1^0 \right|^2  \right)
	+ Y_{\nu}^2 \sum_{i=1}^3 \left| \snuc_i \right|^2 \nonumber \\ 
&~~~~
	+ \lambda^2 \left| \snuc_1 + \snuc_2 + \snuc_3 \right|^2 \\
	\left( M_{H^{\pm}}^2 \right)_{H_1 \tilde{\ell}_{L_i}} &= 
	\left( \frac{1}{2} g_2^2 - Y_{e_i}^2 \right) v_{\snu} \left( H_1^0 \right)^{\ast} 
	- Y_{\nu} \lambda \left( \snuc_i \right)^{\ast} \sum_{i=k}^3 \snuc_k \\
	\left( M_{H^{\pm}}^2 \right)_{H_1 \tilde{\ell}_{R_i}} &= 
	- a_{e} v_{\snu} - Y_{\nu} Y_{e_i} \left( H_2^0 \snuc_i \right)^{\ast} \\
	\left( M_{H^{\pm}}^2 \right)_{H_2 \tilde{\ell}_{L_i}} &= 
	\left( \frac{1}{2} g_2^2 - Y_{\nu}^2 \right) v_{\snu} H_2^0
	+ \lambda Y_{\nu} H_1^0 H_2^0 
	- Y_{\nu} \kappa \left( \snuc_i \right)^2
	- a_{\nu} \left( \snuc_i \right)^{\ast} 
\end{align}
\begin{align}
	\left( M_{H^{\pm}}^2 \right)_{H_2 \tilde{\ell}_{R_i}} &= 
	- Y_{\nu} Y_{e_i} H_1^0 \left( \snuc_i \right)^{\ast}
	- \lambda Y_{e_i} v_{\snu} \sum_{k=1}^3 \left( \snuc_k \right)^{\ast} \\
	\left( M_{H^{\pm}}^2 \right)_{\tilde{\ell}_{L_i} \tilde{\ell}_{L_j}} &= 
	\delta_{ij} \left[ 
	m_{L}^2 
	+ \frac{1}{4} \left( g_1^2 - g_2^2 \right) \left( \left| H_1^0 \right|^2 - \left| H_2^0 \right|^2 + 3 v_{\snu}^2 \right) 
	+ Y_{e_i}^2 \left| H_1^0 \right|^2 
	\right]
	+ \frac{1}{2} g_2^2 v_{\snu}^2 \nonumber \\
&~~~~ + Y_{\nu}^2 \snuc_i \left( \snuc_j \right)^{\ast} \\
	\left( M_{H^{\pm}}^2 \right)_{\tilde{\ell}_{R_i} \tilde{\ell}_{R_j}} &= 
	\delta_{ij} \left[ 
	m_{e^c}^2 
	- \frac{1}{2} g_1^2 \left( \left| H_1^0 \right|^2 - \left| H_2^0 \right|^2 + 3 v_{\snu}^2 \right) 
	+ Y_{e_i}^2 \left| H_1^0 \right|^2
	\right]
	+ Y_{e_i} Y_{e_j} v_{\snu}^2 \\
	\left( M_{H^{\pm}}^2 \right)_{\tilde{\ell}_{L_i} \tilde{\ell}_{R_j}} &= 
	\delta_{ij} \left[ 
	a_{e_i} H_1^0 
	- \lambda Y_{e_i} \left( H_2^0 \right)^{\ast} \sum_{k=1}^3 \left( \snuc_k \right)^{\ast}
	\right]~~.
\end{align}

\textbf{Neutral Scalars.}  The neutral Higgses mix with the left- and
right-handed sneutrinos in a $16 \times 16$ matrix $M_{H_0}^2$.  At
the EWSB vacuum which respects CP, this matrix block diagonalizes into
CP-even and CP-odd sectors.  In order to study the phase transition in
which there are transitional CP-violating phases, we must retain the
off-diagonal blocks.  In the basis $\phi^T = \left\{ \Re H_0, \Im H_0
\right\}$ where $H_0 = \left\{ H_1^0, H_2^0, \snuc_1, \snuc_2,
\snuc_3, \snu_1, \snu_2, \snu_3 \right\}$ the mass term is given by
$\mathcal{L} \ni - \frac{1}{2} \phi^T M_{H_0}^2 \phi$ with $n_{H_0} =
1$.  One can obtain the mass matrix
\begin{align}
	\left( M_{H_0}^2 \right)_{ij} & = 
\begin{pmatrix}
M_{ \Re H_0}^2 & M_{\slashed{CP}}^2 \\
M_{\slashed{CP}}^2 & M_{ \Im H_0}^2 
\end{pmatrix}_{ij} = \frac{\partial^2 \bar{V}_0}{\partial \phi_i \partial \phi_j} .
\end{align}
by differentiating the full scalar potential
\begin{align}\label{eq:V0bar}
	\bar{V}_0 &= V_0 +
	m_L^2 \sum_i \left| \snu_i \right|^2 
	+ Y_{\nu}^2 \left( \left| H_2^0 \right|^2 \sum_i \left| \snu_i \right|^2 + \left| \sum_i \snu_i \snuc_i \right|^2 \right) \nonumber \\
	& + \sum_i \left[ a_{\nu} H_2^0  \snu_i \snuc_i + Y_{\nu} \kappa \left( H_2^0 \snu_i \right)^{\ast} \left( \snuc_i \right)^2 + \hc \right] \nonumber  \\
	&- \lambda Y_{\nu}  \sum_i \left[  \left| H_2^0 \right|^2 H_1^0  \snu_i^{\ast} + H_1^0   \snuc_i   \sum_j \left( \snu_j \snuc_j \right)^{\ast} + \hc \right] \nonumber \\
	& + \left[  \frac{g_1^2 + g_2^2}{8} \left( \left| H_1^0 \right|^2 - \left| H_2^0 \right|^2 + \sum_i \left| \snu_i \right|^2 \right)^2  -  \frac{g_1^2 + g_2^2}{8} \left( \left| H_1^0 \right|^2 - \left| H_2^0 \right|^2 \right)^2 \right]
\end{align}
where the dominant contribution $V_0$ is given by \eref{eq:V0}.

\section{\label{sec:bosonicthermalmasses}Bosonic Thermal Masses}

In order to calculate the daisy resummation \eref{eq:Vdaisy} we
require the thermal mass corrections $\Pi_b$.  For the Higgs and
singlet fields we compute the thermal mass corrections from the
thermal effective potential using the procedure explained in this
section.  For the left-handed sneutrinos we use
\begin{align}
	\frac{\Pi_{\snu_i}}{T^2} = \frac{g_1^2}{8} + \frac{7 g_2^2}{24} + \frac{5 Y_{e_i}^2}{24} + \frac{Y_{\nu}^2}{4} 
\end{align}
which can be calculated by assuming that all species that are summed
in $\Delta V_1^T$ are light.  For the remaining bosonic species, we
use the thermal mass functions calculated for the nMSSM by
\cite{Menon:2004wv} in which the authors assumed that the Higgs,
Higgsinos, electroweak gauginos, and SM particles were light.

First, we evaluate the thermal effective potential correction
\eref{eq:DV1T} as a function of the eigenvalues of the field-dependent
mass matrices listed in Appendix \ref{sec:massmatrices}.  Let
$\tilde{m}^2_{ij}$ be the $j^{\mathrm{th}}$ eigenvalue of the
$i^{\mathrm{th}}$ mass matrix with has $n_i$ associated degrees of
freedom.  By writing the traces as a sum over eigenvalues and using that 
$n_i < 0$ for fermionic species, \eref{eq:DV1T} can be written as
\begin{align}
	\Delta V_1^T = \frac{T^4}{2 \pi^2} \sum_i \left| n_i  \right|
	\begin{cases}
	\sum_j J_{B} \left( \tilde{m}^2_{ij} / T^2 \right) & \text{i bosonic} \\
	- \sum_j J_{F} \left( \tilde{m}^2_{ij} / T^2 \right) & \text{i fermionic} \\
	\end{cases} .
\label{eq:wherenifirstused}
\end{align}
In the high-temperature limit, $\tilde{m}^2_{ij} \ll T^2$ the bosonic and fermionic thermal functions can be expanded as 
\begin{align}
	J_B \left( y \right) \xrightarrow{y \ll 1} \frac{\pi^2}{12} y + \ord{y^{3/2}} \\
	J_F \left( y \right) \xrightarrow{y \ll 1} - \frac{\pi^2}{24} y + \ord{y^2}
\end{align}
plus field independent terms.  Second, we define the high-temperature thermal potential correction by imposing a sharp cutoff at $\tilde{m}_{ij}^2 = 2 T^2$ to obtain
\begin{align}\label{eq:DV1Thigh}
	\Delta V_1^{T, \mathrm{high}}  = \frac{1}{48} T^2 \sum_i \left| n_i \right|  
	\begin{cases}
	\sum_j 2 \tilde{m}_{ij}^2 & \text{i bosonic} \\
	\sum_j \tilde{m}_{ij}^2 & \text{i fermionic} \\
	0 & \tilde{m}_{ij}^2 < 2 T^2 \\
	\end{cases}
\end{align}
Third, we extract the thermal mass corrections by differentiating with respect to the Higgs and singlet fields,
\begin{align}\label{eq:ThermalMassCorrection}
	\Pi_{\phi_i} = T^2 \left[ \frac{\partial^2}{\partial \phi_i^2} \frac{ \Delta V_1^{T, \mathrm{high}}}{T^2} \right]_{H_1^0 = H_2^0 = \snuc_j = 0, T = 100 \GeV}
\end{align}
where $\phi_i \in \left\{ H_1^0, H_2^0, \snuc_k \right\}$.  The
derivatives are evaluated at the origin in field space such that
$\Pi_{\phi_i}$ is accurate in the high-temperature vacuum.  Because the
derivative in \eref{eq:ThermalMassCorrection} has only weak field
dependence, we expect this expression for $\Pi_{\phi_i}$ to be
accurate even for our multistep phase transitions in which the
singlets have VEVs before the EWPT.  The value of $T$ used in
\eref{eq:ThermalMassCorrection} only affects the location of the
cutoff in \eref{eq:DV1Thigh}.  We have chosen the temperature $T = 100
\GeV$ to be at the appropriate scale for our phase transitions and
such that $\Pi_{\phi_i}$ does not vary discontinuously in the region
of parameter space with first order phase transitions.  Using this
procedure we obtain $\Pi_{H_1^0} \approx 0.11~T^2, \Pi_{H_2^0} \approx
0.40~T^2, \Pi_{\snuc_i} \approx 0.20~T^2$ over the region of parameter
space with phase transitions.

\section{\label{sec:derivingboundaries}Analytic Derivation of Parameter Space Boundaries}

The boundaries in Fig.  \ref{fig:Scan1} can be understood
analytically.  In this section, we derive expressions for each of
the boundaries and discuss the parametric dependence.

At the interface of regions I and II, the electroweak vacuum develops
a tachyonic direction at tree level and $\det M_{\Re H_0}^2 = 0$.
Since $M_{\Re H_0}^2$ is an 8 by 8 matrix, it would not be useful to
write out its determinant.  Instead, we observe that the tachyonic
direction is directed along $
\{ H_1^0 / v_1,  H_2^0 / v_2, \snuc_i
  / v_{\snuc} \approx 1, \snu_i = 0 \}$.

At the boundary between region IIa and III, the minima at $\x_{012}$ and $\y_{000}$ are degenerate at one-loop.  Note that this degeneracy cannot occur at tree level.  To see why, write 
\begin{align}
	V_0 \left( \x_{012} \right) - V_0 \left( \y_{000} \right) = \Delta V_0^a + \Delta V_0^b
\end{align}
with $\Delta V_0^a = V_0 \left( \x_{012} \right) - V_0 \left( \x_{000}
\right)$ and $\Delta V_0^b = V_0 \left( \x_{000} \right) - V_0 \left(
\y_{000} \right)$.  The tree level $\left( \mathbb{Z}_3 \right)^3$
symmetry ensures $\Delta V_0^a = 0$.  Using the minimization
equations \eref{eq:MinEqns} we can write
\begin{align}
	\Delta V_0^b = \frac{1}{8} \left[ \left( g_1^2 + g_2^2 \right)  \cos^2 2 \beta + 6 \lambda^2  \sin^2 2 \beta \right] v^4 > 0 
\end{align}
where we have also neglected terms suppressed by $v_{\snu} / v$ and $Y_{\snu}$.  
At one-loop order we calculate the difference in the potential as
\begin{align}
	V_1^0 \left( \x_{012} \right) - V_1^0 \left( \y_{000} \right) = \Delta V_1^a + \Delta V_1^b 
\end{align}
where $V_1^0$ is the one-loop, zero temperature effective potential and $\Delta V_1^{a,b}$ are defined analogously as above.  We expect $\Delta V_1^a$ to be nonzero and sensitive to the radiative corrections because the $\left( \mathbb{Z}_3 \right)^3$ symmetry is broken to $\mathbb{Z}_3$.  The terms responsible for this symmetry breaking are the superpotential term $W \ni \lambda \hat{H}_1^0 \hat{H}_2^0 \hat{\nu}^c_i$ and corresponding A-term in the soft SUSY-breaking Lagrangian.  We calculate 
\begin{align}\label{eq:DV1a}
	64 \pi^2 \Delta V_1^a & = 
	 6 m_{\ch}^4 \log \frac{m_{\ch}^2}{e^{3/2} \mu^2} 
	+ 4 \left[m_{H_1}^4 \log \frac{m_{H_1}^2}{e^{3/2} \mu^2} + m_{H_2}^4 \log \frac{m_{H_2}^2}{e^{3/2} \mu^2}  -  \sum_{\pm} m^4_{\pm} \log \frac{m^2_{\pm}}{e^{3/2} \mu^2} \right] \nonumber \\
	m_{\pm}^2 & = m_{\ch}^2 + \frac{m_{H_1}^2 + m_{H_2}^2}{2} \pm \frac{1}{2} \sqrt{ \left( m_{H_1}^2 - m_{H_2}^2 \right)^2 + \sigma^2 m_{\ch}^4 }
\end{align}
For the sake of discussion, we can approximate the logarithms in the second term as order one numbers and eliminate the soft masses using \eref{eq:MinEqns} to obtain
\begin{align}\label{eq:DV1aApprox}
	64 \pi^2 \Delta V_1^a \approx 
	6 m_{\ch}^4 \log \frac{m_{\ch}^2}{e^{3/2} \mu^2} 
	-2 m_{\ch}^4 \left(  \sigma^2 + 4 \sigma \csc 2 \beta -4 \right)
	+ 24 \lambda^2 v^2 m_{\ch}^2
\end{align}
Since we are simply trying to estimate the parametric dependence, we can approximate $\Delta V_1^b \approx \Delta V_0^b$.  By requiring that the minimum at $\x_{012}$ is not deeper than the EWSB vacuum, we obtain the bound $\Delta V_1^a + \Delta V_0^b \geq 0$, which is saturated at the interface of regions IIa and III.  This bound disfavors large $m_{\ch}$ and large $\sigma$ because of the $- m_{\ch}^4 \sigma^2$ term in $\Delta V_1^a$.  

At the boundary where regions III and IIb meet, the EWSB is degenerate with the origin in field space at one-loop.  Neglecting the radiative corrections we can approximate the splitting as $V_1^0 \left( \x_{O} \right) - V_1^0 \left( \y_{000} \right) \approx V_0 \left( \x_{O} \right) - V_0 \left( \y_{000} \right) \equiv \Delta V_0^c$ with 
\begin{align}
	\Delta V_0^c &= \frac{1}{8} \left[ \left( g_1^2 + g_2^2 \right) \cos^2 2 \beta + 6 \lambda^2  \sin^2 2 \beta \right] v^4 \nonumber \\
	& + m_{\ch}^2 v^2 \left[ 1 - \frac{\kappa \sin 2 \beta}{6 \lambda} - \frac{\sigma \sin 2 \beta}{4} \right] + \frac{m_{\ch}^4 \kappa^2}{27 \lambda^4} + \frac{a_{\kappa} m_{\ch}^3}{27 \lambda^3}
\end{align}
To prevent the origin from becoming the global minimum we require $\Delta V_0^c > 0$ which favors larger $m_{\ch}$ and smaller $\sigma$.  

At the boundary between regions IIc and III, the one-loop potential has degenerate minima at $\y_{012}$ and $\y_{000}$.  We can compute the splitting $\Delta V_0^d \equiv V_0 \left( \y_{012} \right) - V_0 \left( \y_{000} \right)$ by neglecting the radiative corrections to find
\begin{align}\label{eq:DV0d}
	\Delta V_0^d = \frac{1}{2} m_{\ch}^2 v^2 \left( \sigma \sin 2 \beta - 2 \right) .
\end{align}
The condition that the EWSB minimum at $\y_{000}$ is absolutely stable requires $\Delta V_0^d > 0$ which imposes the lower bound $\sigma \gtrsim 2 \csc 2 \beta \approx 3$ for $\tan \beta = 2.6$.  Figure \ref{fig:Scan1} shows that the IIc-III boundary also depends on $m_{\ch}$ contrary to \eref{eq:DV0d}, but this is a result of the radiative corrections.

\section{\label{sec:slection} Selecting a CP-Even Vacuum}

In this appendix, we show formally how a superpotential contribution
$\Delta W$ that breaks $\mathbb{Z}_3$ weakly can be constructed to make
the CP conserving vacuum to have the lowest energy perturbatively in
the absence of any explicit CP violating parameters.  Consider the
superpotential\begin{equation} W=W_{0}+\Delta W\end{equation} where
$\Delta W$ represents a irrelevant operator perturbation to
renormalizable $W_{0}$. We then have\begin{eqnarray}
\sum_{i}\left|\frac{\partial W}{\partial\phi_{i}}\right|^{2} & = &
\sum_{i}\left\{ \left|\frac{\partial
  W_{0}}{\partial\phi_{i}}\right|^{2}+2\Re\left[\left(\frac{\partial\Delta
    W}{\partial\phi_{i}}\right)\left(\frac{\partial
    W_{0}}{\partial\phi_{i}}\right)^{*}\right]\right\} \\ & = &
V_{0}+\sum_{i}\Delta V_{i}\end{eqnarray} where\begin{equation}
V_{0}\equiv\sum_{i}\left|\frac{\partial
  W_{0}}{\partial\phi_{i}}\right|^{2}\end{equation}
and\begin{equation} \Delta
V_{i}\equiv2\Re\left[\left(\frac{\partial\Delta
    W}{\partial\phi_{i}}\right)\left(\frac{\partial
    W_{0}}{\partial\phi_{i}}\right)^{*}\right]\end{equation} to
leading order in $\Delta W$ which we will call $\mathcal{O}(\delta)$.
As usual, $W_{0}$ and $\Delta W$ are holomorphic polynomials in
fields. Considering $\phi_{i}\rightarrow\phi_{i}^{*}$ as a representation of
$\mathbb{Z}_{2}$ which we will call 2, we have \begin{equation}
  \left(\frac{\partial\Delta
    W}{\partial\phi_{i}}\right)\left(\frac{\partial
    W_{0}}{\partial\phi_{i}}\right)^{*}\end{equation} being a representation
of \begin{equation} \oplus\sum_{u}(2^{u}\otimes\bar{2}^{2})\equiv
  R.\end{equation} If we assume all the coefficients of $W$ and
$\Delta W$ are real, we can write \begin{equation}
  2\Re\left[\left(\frac{\partial\Delta
      W}{\partial\phi_{i}}\right)\left(\frac{\partial
      W_{0}}{\partial\phi_{i}}\right)^{*}\right]=R\oplus\mathbb{Z}_{2}(R)\end{equation}
where $\mathbb{Z}_{2}(\mathbb{Z}_{2}(R))=R.$ Hence, we see that
$\Delta V_{i}$ is a singlet under $\mathbb{Z}_{2}$. Given that $\Delta
V_{i}$ is a polynomial in $a_{j}\equiv\Re\phi_{j}$ and
$b_{j}\equiv\Im\phi_{j}$ and since under
$\mathbb{Z}_{2}:\{a_{j}\rightarrow a_{j},b_{j}\rightarrow-b_{j}\}$, we
must have \begin{equation} \Delta
  V_{i}=\sum_{k}\sum_{m}c_{km}^{i}P_{k}(\{a_{j}\})S_{m}(\{b_{j}\})\label{eq:basisexpansion}\end{equation}
where $S_{m}$ represents a basis of $\mathbb{Z}_{2}$ singlet
polynomial functions composed of $b_{j}$ and $P_{k}$ is a basis of
polynomial functions composed of $a_{j}$. Note that here
$c_{km}^{i}=\mathcal{O}(\delta).$ Hence, given that the part of the
effective potential not associated with $\Delta W$ had a minimum at
$\vec{\phi}=\vec{v}(s)$ where $s\in\{-1,0,1\}$ parametrizes the
$\mathbb{Z}_{3}$ fundamental representation elements and $b_{j}|_{\vec{v}(0)}=0$
is the singlet element, the energy shift due to $\Delta W$ to
$\mathcal{O}(\delta)$ is\begin{equation}
\Delta\rho(s)\equiv\sum_{i}\Delta
V_{i}|_{\vec{v}(s)}=\sum_{i}\sum_{k}\sum_{m}c_{km}^{i}P_{k}(\{v_{j}\cos\left(\frac{s2\pi}{3}\right)\})S_{m}(\{v_{j}\sin\left(\frac{s2\pi}{3}\right)\}),\end{equation}
Note that $\Delta\rho(1)=\Delta\rho(-1)$. Hence, we only need to
determine whether $\Delta\rho(1)-\Delta\rho(0)>0$ to see if CP
singlet has the lowest energy. Since
$c_{km}^{i}\propto\mbox{sgn}\Delta W$, we can simply flip the sign of
$\Delta\rho(1)-\Delta\rho(0)$ by flipping the sign of $\Delta W$ if
the original choice of sign gives $\Delta\rho(1)-\Delta\rho(0)<0$.  Of
course, all of this is under the assumption that the potential is not
destabilized by the nonrenormalizable operators such that the
smallness of the perturbation order $\delta$ is meaningful. Stability
is generic if the nonrenormalizable terms are dominated by the
perturbations in the superpotential since the superpotential
contribution is positive definite.

\providecommand{\href}[2]{#2}\begingroup\raggedright\endgroup

\end{document}